\documentclass[traditabstract]{aa}
\usepackage[dvips]{graphicx}
\usepackage{txfonts}
\usepackage{natbib}
\bibpunct{(}{)}{;}{a}{}{,}

\begin{document}

\title{Making CMB temperature and polarization maps
with Madam}

\author
{E. Keih\"anen$^1$\thanks{E-mail: elina.keihanen@helsinki.fi} \and
 R. Keskitalo\inst{1,2} \and H. Kurki-Suonio\inst{1,2} \and T. Poutanen\inst{1,2,3}
 \and A.-S. Sirvi\"o\inst{1} }

\institute{
University of Helsinki, Department of Physics,
P.O. Box 64, FIN-00014, Helsinki, Finland \and
Helsinki Institute of Physics, P.O. Box 64, FIN-00014, Helsinki, Finland
 \and
 Mets\"{a}hovi Radio Observatory, Helsinki University of Technology,
 Mets\"{a}hovintie 114, FIN-02540 Kylm\"{a}l\"{a}, Finland
}

\abstract{ {\small MADAM} is a CMB map-making code, designed to make
temperature and polarization maps of time-ordered data of total
power
experiments like {\sc Planck}. The algorithm is based on the destriping
technique, but it also makes use of known noise properties in the
form of a noise prior. The method in its early form was presented in
an earlier work by Keih\"{a}nen et al.~(2005).
In this paper we present an update of the method, extended to
non-averaged data, and include polarization. In this method the
baseline length is a freely adjustable parameter, and destriping can
be performed at a different map resolution than that of the final
maps. We show results obtained with simulated data. This study is
related to {\sc Planck} LFI activities.}


\maketitle

%
%

\section{Introduction}

{\small MADAM} is a map-making code, designed to build full-sky maps of
the cosmic microwave background (CMB) anisotropy. The code was developed
for the {\sc Planck} satellite, but is applicable to any total power
CMB experiment. The code takes as input the
time-ordered data (TOD) stream, together with pointing information,
and produces full-sky maps of CMB temperature and polarization.

{\sc Planck} data is contaminated by slowly-varying $1/f$ noise, which
must be removed in the map-making process. The {\small MADAM} algorithm is
based on the destriping technique \citep{Burigana1997,
Delabrouille1998, Maino1999,Maino2002, Keihanen2004}, where the
correlated noise component is modelled by a linear combination of
some base functions. Unlike conventional destriping, {\small MADAM} also
makes use of a priori information on the noise spectrum.
A similar method has recently been implemented by \cite{Sutton09}.

The {\small MADAM}\ map-making algorithm was first introduced by
\cite{Keihanen2005}. Since then, the code has undergone significant
development. In the first paper, we applied the code to coadded
data, where data was averaged over the 60 scanning circles of the
same ``ring'', i.e., the data segment between two satellite spin
repointings, to reduce the data volume. The 60 scanning circles were
assumed to fall exactly on top of each other. Improved computational
resources now allow us to apply the method to non-averaged data.
Averaging the data reduced the data volume by a large factor. On the
other hand, not averaging the data simplifies the analysis. It also
lifts one non-realistic simplification, since the subsequent
scanning circles do no coincide exactly in reality.

In the first paper we considered only total intensity maps.
In this study we also consider polarization.

When dealing with ring-averaged data, it was natural to use the
scanning ring length as a basic length unit. In the first paper we
fitted different base functions (Fourier components, Legendre
functions) to rings. In this study we abandon the concept of "ring",
and make no assumptions on the scanning pattern. In this approach it
is a natural choice to fit the data with uniform baselines only, but
to vary their length. This also simplifies the method.

Though the algorithm makes no assumptions on the scanning pattern,
the actual pattern will have an impact on the choice of the optimal baseline length.

%
%
\begin{figure}
  \includegraphics[width=9cm]{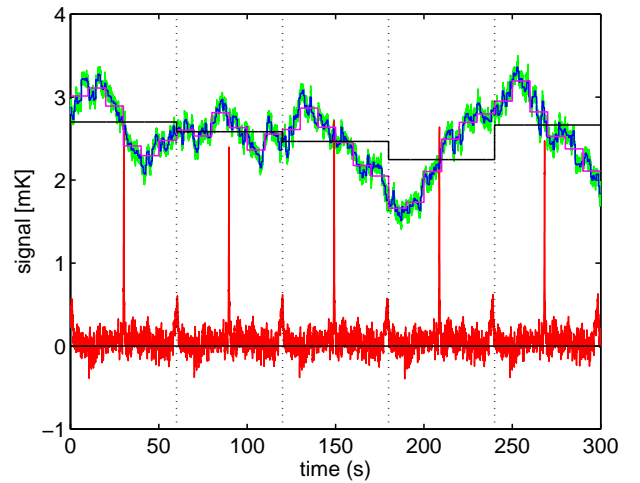}
\caption{
{\bf Time-ordered data}. 
We show a 5-minute excerpt of
simulated $1/f$ noise, fitted with 1 min, 10 s, or 0.625 s baselines
({\it black, purple, blue}, respectively). We show also the signal TOD ({\it red}). 
The vertical dotted lines mark the beginning
and end of 1 min scanning periods. }
 \label{fig:todcomp}
\end{figure}
%

We plot in Fig.~\ref{fig:todcomp} a 5-minute excerpt of simulated
$1/f$ noise. We show also 1 min, 10 s, and 0.625 s  baselines fitted
to the noise stream. Short baselines naturally follow the actual
noise stream more closely. We show in the same figure also the
signal TOD. 

The signal shows a clear periodicity, following the 1
min scanning period. The strong peaks in the signal are the galaxy.
The signal plotted here includes CMB and foreground, but no dipole.
The total observed TOD is a sum of the signal,
$1/f$ noise with a knee frequency of 50 mHz, and white noise with
a  standard deviation of $\sigma=2.7$ mK. We do
not plot the white noise component, which would obscure the figure.

In this paper we present results obtained with simulated data.
We demonstrate how changing various input parameters in {\small MADAM}
affects the quality of the output maps.
The most important input parameters are the baseline length and
the destriping resolution.

Since the purpose of this study is to present one map-making method,
rather than to demonstrate the performance of the {\sc Planck} experiment,
we have ignored some systematic effects that are present  in
realistic data, but which {\small MADAM} does not attempt to correct for.
These effects include pointing errors, asymmetric beams and finite
integration time. Their impact on the accuracy of CMB maps has been
discussed in several papers
\citep{Poutanen2006,Ashdown2007a,Ashdown2007b,Ashdown2009}.

%
%

\section{Map-making problem}
\label{sec:ml}

In the following we go through the maximum-likelihood analysis on
which our map-making method is based. Essentially the same analysis
was presented by \cite{Keihanen2005}, but we repeat it here for the
sake of self-consistency, in a somewhat more concise form. We widen
the interpretation of various matrices to include polarization, and
extend the results to a case where the final map is constructed at a
resolution different from the destriping resolution.

We write the time-ordered data (TOD) stream as
 \begin{equation}
    \vec{y} = \mathbf{P}\vec{m} +\vec{n}'. \label{tod1}
\end{equation}
Here the first term presents the CMB signal and the second
term presents noise.
Vector $\vec{m}$ presents the pixelized CMB+foreground map, and pointing
matrix $\mathbf{P}$ spreads it into TOD.
$\mathbf{P}$ is a matrix of size ($N_\mathrm{t}$,$3N_\mathrm{pix}$),
where $N_\mathrm{t}$ is the length of the TOD vector.
If the analysis includes several detectors,
they can be thought as catenated into one vector.

Since we are including polarization, map $\vec{m}$ is an object of $3N_\mathrm{pix}$
elements, where $N_\mathrm{pix}$ is the number of sky pixels.
In each pixel, the map consists of three values, corresponding to the three
Stokes parameters I,Q,U. The observed signal can be written as
\begin{equation}
     y_t = I_t +Q_t\cos(2\psi_t) +U_t\sin(2\psi_t).  \label{poltod}
\end{equation}
Here $\psi$ is the orientation of polarization sensitivity, dependent both
on the momentary orientation
of the spacecraft, and on the detector's orientation on the focal plane.
The factors $1,\cos(2\psi_t),\sin(2\psi_t)$ constitute the pointing matrix $\mathbf{P}$.
If the detector is not polarization sensitive, the cosine and sine terms are dropped.

We divide the noise contribution into a correlated noise component
and white noise, and model the correlated part as a sequence of uniform baselines,
\begin{equation}
    \vec{n}' = \mathbf{F}\vec{a} +\vec{n}.   \label{noisetod}
\end{equation}
Vector $\vec a$ contains the unknown amplitudes of the baselines,
and matrix $\mathbf{F}$ spreads them into TOD.
Matrix $\mathbf{F}$ consists of zeroes and ones, indicating which samples
belong to which baseline.

Assuming that the white noise
component and the correlated noise component are independent,
the total noise covariance in time domain is given by
\begin{equation}
   \mathbf{C}_\mathrm{t}= \langle\vec n'(\vec n')^T\rangle
       = \mathbf{F}\mathbf{C}_\mathrm{a}\mathbf{F}^T +\mathbf{C}_\mathrm{n} \label{totcov}
\end{equation}
where $\mathbf{C}_\mathrm{n}=\langle\vec n\vec n^T\rangle$ is the white noise
covariance, $\mathbf{C}_\mathrm{a}=\langle\vec a\vec a^T\rangle$ is the covariance
matrix for the baseline amplitudes $\vec{a}$, and $\langle x\rangle$
denotes the expectation value of quantity $x$.

Maximum likelihood analysis yields the chi-square
minimization function
\begin{eqnarray}
      \chi^2 &=&(\vec{y}-\mathbf{F}\vec{a}-\mathbf{P}\vec{m})^T
                \mathbf{C}_\mathrm{n}^{-1}(\vec{y}-\mathbf{F}\vec{a}-\mathbf{P}\vec{m})\nonumber\\
             && +\vec{a}^T\mathbf{C}_\mathrm{a}^{-1}\vec{a}.
      \label{chi1}
\end{eqnarray}

We want to minimize $(\ref{chi1})$ with respect to both $\vec{a}$ and
 $\vec{m}$.
Minimization with respect to $\vec{m}$ gives
\begin{equation}
          \vec{m}= (\mathbf{P}^T\mathbf{C}_\mathrm{n}^{-1}\mathbf{P})^{-1}
           \mathbf{P}^T\mathbf{C}_\mathrm{n}^{-1}(\vec{y}-\mathbf{F}\vec{a}) \, .  \label{mlmap1}
\end{equation}
Substituting Eq.~(\ref{mlmap1}) back into Eq.~(\ref{chi1}) and minimizing with respect to $\vec{a}$,
we obtain an estimate for the amplitude vector $\vec a$.
The solution is given by
\begin{equation}
    (\mathbf{F}^T\mathbf{C}_\mathrm{n}^{-1}\mathbf{Z}\mathbf{F}+\mathbf{C}_\mathrm{a}^{-1})\vec{a}
     = \mathbf{F}^T\mathbf{C}_\mathrm{n}^{-1}\mathbf{Z}\vec{y}  \label{linc}
\end{equation}
where
\begin{equation}
      \mathbf{Z} = \mathbf{I}-\mathbf{P}(\mathbf{P}^T\mathbf{C}_\mathrm{n}^{-1}\mathbf{P})^{-1} 
      \mathbf{P}^T\mathbf{C}_\mathrm{n}^{-1} \, .
 \label{zdef}
\end{equation}

A similar analysis without the noise covariance term $\mathbf{C}_\mathrm{a}$
was presented by \cite{HX2009},
along with an extensive discussion of the properties of the various matrices involved.

In some situations it is beneficial to solve the baselines at a resolution different
from the resolution of the final map. Particularly,  a strong {\it signal error} caused by
a strong foreground signal can be reduced by destriping at a high resolution.
Signal error is discussed in Sect. \ref{sec:residualerror}.
We now extend our analysis to a case where the two resolutions are different.

Matrix $\mathbf{P}$ depends on resolution. We define two different pointing
matrices: matrix $\mathbf{P}_\mathrm{c}$ is constructed at the {\it destriping resolution}
and matrix $\mathbf{P}_\mathrm{m}$ at the {\it map resolution}.

We estimate the baseline amplitudes by solving vector $\vec a$  from Eq. (\ref{linc}),
where now
\begin{equation}
      \mathbf{Z} = \mathbf{I}-\mathbf{P}_\mathrm{c}(\mathbf{P}^T\mathbf{C}_\mathrm{n}^{-1}\mathbf{P}_\mathrm{c})^{-1}
              \mathbf{P}_\mathrm{c}^T\mathbf{C}_\mathrm{n}^{-1} \, .\label{zref}
\end{equation}
The final map is then constructed by binning the cleaned TOD into a map at a different
resolution as
\begin{equation}
          \vec{m}= (\mathbf{P}_\mathrm{m}^T\mathbf{C}_\mathrm{n}^{-1}\mathbf{P}_\mathrm{m})^{-1}
           \mathbf{P}_\mathrm{m}^T\mathbf{C}_\mathrm{n}^{-1}(\vec{y}-\mathbf{F}\vec{a}) \, .  \label{mlmap}
\end{equation}
Strictly speaking, separating the two pointing matrices breaks the maximum
likelihood analysis presented above, in the sense that the map constructed through
Eqs. (\ref{linc}), (\ref{zref}), and (\ref{mlmap}) is not the solution of any maximum likelihood problem.
In practice, the division works well and is intuitive.

We define two other maps, which are useful for further analysis,
and are products of the {\small MADAM} code.
The {\it binned map} is constructed without baseline extraction as
\begin{equation}
   (\mathbf{P}_\mathrm{m}^T\mathbf{C}_\mathrm{n}^{-1}\mathbf{P}_\mathrm{m})^{-1}
   \mathbf{P}_\mathrm{m}^T\mathbf{C}_\mathrm{n}^{-1}\vec y,  \label{binmap}
\end{equation}
and is useful in signal-only simulations.
The {\it sum map}
\begin{equation}
    \mathbf{P}_\mathrm{m}^T\mathbf{C}_\mathrm{n}^{-1}(\vec y-\mathbf{F}\vec a)
\end{equation}
is useful in the case of incomplete sky coverage,
since it is well-defined also in pixels with poor sampling
of polarization directions.

%
%

\section{Noise prior}
\label{sec:noiseprior}

We consider now the statistical properties of the correlated noise component,
assumed to be stationary, and derive a formula for the {\it noise prior} $\mathbf{C}_\mathrm{a}$.
In the first paper \citep{Keihanen2005} we presented a method for the computation
of the noise prior for a general baseline function.
In the case of uniform baselines, the covariance can be constructed by a simpler
method as follows.

Let $c(t)$ denote the auto-covariance function of the time-ordered noise stream.
That is, the expectation value of the product of two samples of the noise stream,
time $t$ apart, is given by
\begin{equation}
    c(t) = \langle y(t_0)y(t_0+t)\rangle.
\end{equation}
The auto-covariance may be expressed as a Fourier transform
of the noise spectrum as
\begin{equation}
    c(t) = \int_{-\infty}^{\infty} P(f)e^{i2\pi ft}df
\end{equation}
This sets the normalization convention applied in this paper.

We assume that the spectrum $P(f)$ is known.
We study now the covariance between two baselines of length $T$.
We define the {\it reference value} for a baseline offset, extending from time $t$ to time $t+T$,
as an average of the noise stream,
\begin{equation}
   \hat a(t) = \frac1T \int_t^{t+T}y(t')dt'.
\end{equation}
We calculate the noise prior $\mathbf{C}_\mathrm{a}$ as the covariance matrix
between these
reference values.

Consider a sequence of baselines, starting at times $t=t_0,t=t_0+T,t=t_0+2T\ldots$
The correlation between the two baselines only depends on their distance.
We denote by $c_\mathrm{a}(k)$ any of the elements of $\mathbf{C}_\mathrm{a}$ on the $k$:th diagonal.
We may write
\begin{equation}
   c_\mathrm{a}(k) = \langle \hat a(t)\hat a(t+kT)\rangle.
\end{equation}
We need the inverse of matrix $\mathbf{C}_\mathrm{a}$, which appears in the
destriping equation (\ref{linc}). The inverse is computed most
effectively by the Fourier technique. We thus actually need the
spectrum of the reference baselines, defined as 
\begin{equation}
    P_\mathrm{a}(f) = \sum_{k=-\infty}^{\infty} c_\mathrm{a}(k) e^{-i2\pi kfT} .
\end{equation}

We now want to find a relation between $P_\mathrm{a}(f)$ and $P(f)$.
Using the definition of $\hat a$ we write
\begin{equation}
   c_\mathrm{a}(k) = \frac1{T^2}\int_0^T dt \int_0^T dt'
            \langle y(t_0+t)y(t_0+kT+t')\rangle,
\end{equation}
which can be written in terms of the auto-covariance function $c(t)$ as
\begin{equation}
  c_\mathrm{a}(k) = \frac1{T^2}\int_0^T \int_0^T dt dt' c(kT+t'-t).
\end{equation}
We further express this in terms of the noise spectrum,
\begin{equation}
  c_\mathrm{a}(k) = \frac1{T^2}\int_0^T \int_0^T dt dt'
            \int_{-\infty}^{\infty} P(f)e^{i2\pi f(kT+t'-t)}df.
\end{equation}
The integrals over $t$  and $t'$ can be carried out analytically,
yielding
\begin{equation}
  c_\mathrm{a}(k) = \int_{-\infty}^{\infty} \!\! P(f)e^{i2\pi kfT}df
            \frac1{(2\pi fT)^2} (2-2\cos(2\pi fT)).
\end{equation}
We can now write out the baseline spectrum as
\begin{equation}
    P_\mathrm{a}(f) = \!\! \sum_{k=-\infty}^{\infty}  \!
             \int_{-\infty}^{\infty} \!\!\! P(f')e^{i2\pi k(f'-f)T}df'
            \frac{2-2\cos(2\pi f'T)}{(2\pi f'T)^2}.   \label{Paint}
\end{equation}
The sum over $k$
yields zero unless $f-f'$ is a multiple of $1/T$,
in which case it gives infinity. We may write, in terms of the delta function,
\begin{equation}
   \sum_{k=-\infty}^{\infty}
             e^{i2\pi k(f'-f)T} = \frac1T\sum_m[\delta(f'-f-m/T)].
\end{equation}
Substituting this into (\ref{Paint}) gives
\begin{equation}
    P_\mathrm{a}(f) = \frac1T \sum_{m=-\infty}^{\infty}
             P(f+m/T) \frac{ (2-2\cos(2\pi fT))}{(2\pi (fT+m))^2}
\end{equation}
To put this into a slightly more elegant form we use the formula
$\cos(2x)=1-2\sin^2x$ and define
\begin{equation}
    g(x) = \frac1{(\pi x)^2} \sin^2(\pi x).
\end{equation}
The spectrum of the reference baselines can now be written in the final form as
\begin{equation}
    P_\mathrm{a}(f) = \frac1T \sum_{m=-\infty}^{\infty}
             P(f+m/T) g(fT+m).  \label{Paspec}
\end{equation}
If the noise spectrum decreases steeply with frequency, as is the case with $1/f$ noise,
it is enough to evaluate only a few terms around $m=0$.

Once the spectrum (\ref{Paspec}) is known, it is straightforward to
compute the term $\mathbf{C}_\mathrm{a}\vec a$, needed in the evaluation of Eq.
(\ref{linc}), for an arbitrary baseline vector $\vec a$, using the
Fourier technique.

%
%

\section{Implementation}

%
%
\begin{figure}
    \includegraphics[width=9cm]{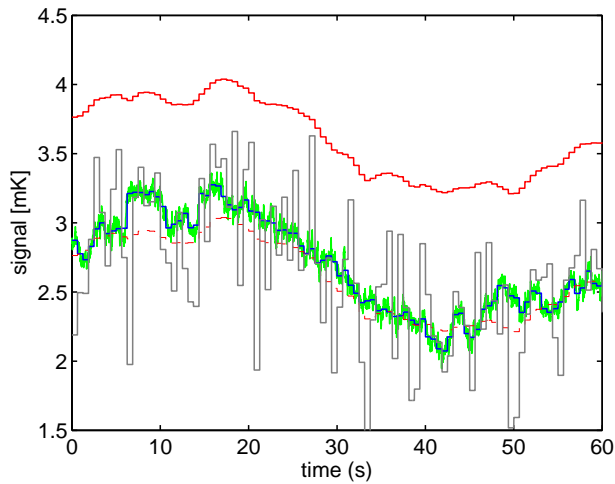}
\caption{
{\bf Recovered $1/f$ noise.}
We show a one minute excerpt of simulated $1/f$ noise ({\it green}),
fitted with 0.625 s baselines (48 samples).
The {\it blue} curve shows the reference baselines.
The {\it gray} curve shows the combined
$1/f$+white noise timeline, averaged over the baseline period.
The {\it red} curve above
shows the baselines recovered by {\small MADAM}.
There is an offset between recovered and actual noise,
due to the fact that destriping cannot determine the mean of the noise stream.
To help the comparison, we show by {\it red dashed} line the recovered noise stream
lowered by 1 mK.
}
\label{fig:todfit}
\end{figure}


{\small MADAM}\ is written in Fortran-90 and parallelized by MPI.

The code takes as an input the time-ordered data stream (TOD) for one
or several detectors, and pointing information, which consists of
three pointing angles $\theta,\phi,\psi$ for each TOD sample. Angles
$\theta$ and $\phi$ define a point on the celestial sphere, while $\psi$
defines the orientation of the polarization sensitivity. Alternatively,
{\small MADAM}\ may construct the pointing angles from satellite pointing
data, using the known focal plane geometry.

The noise prior is constructed from a noise spectrum given as input.
{\small MADAM}\ does not make noise estimation herself.
If the noise spectrum is unknown, the user may optionally turn off the noise prior,
in which case {\small MADAM}\ becomes a traditional destriper.

As output, the code produces full-sky maps of intensity and Q and U polarization.
As additional information, the code may be set to output  matrix
 ($\mathbf{P_\mathrm{m}}^T\mathbf{C}_\mathrm{n}^{-1}\mathbf{P_\mathrm{m}}$),
 a hit count map,  a sum map and a binned map,
 and in case of incomplete sky coverage, a sky mask.
 The code can also store the solved baseline offsets.

\subsection{Solving the baselines}

Matrices $\mathbf{F}$ and $\mathbf{P}$, appearing in the destriping equation
(\ref{linc}), are very large, but sparse. {\small MADAM}\ solves the
baseline offsets from Eq. (\ref{linc}) by the conjugate gradient
technique (CG), which is a standard technique for solving sparse
linear systems, (see, for instance, \cite{NumRes}). {\small MADAM}\ stores
only the non-zero elements of the sparse matrices, and performs the
matrix manipulations algorithmically.

The required number of iteration steps typically varies in the range
of 15-100, depending on the baseline length and other input parameters.
Once the baselines are solved, the final map is constructed by a
binning operation defined by Eq. (\ref{mlmap}). We refer to these
computation phases as the {\it destriping phase} and the {\it
binning phase}.

We show in Fig.~\ref{fig:todfit} a one-minute excerpt of simulated
$1/f$\ noise, together with the corresponding component recovered by
{\small MADAM}. The recovered stream consists of baseline amplitudes
solved by the code. 
We show also the reference baselines, obtained by averaging
the $1/f$ noise over 0.625 s periods.
The reference baselines can be regarded as the goal of baseline determination.

There is a global offset between the recovered and actual
$1/f$\ noise. This is due to the fact that destriping cannot
distinguish between a global offset in the observed signal due to
$1/f$\ noise and a similar offset due to a CMB monopole. In
conventional destriping \citep{Burigana1997, Delabrouille1998,
Maino1999,Maino2002, Keihanen2004, HX2009} this manifests as a zero
eigenvalue of the linear system to be solved. When a noise prior is
present, the linear system is mathematically well-conditioned, 
and all eigenvalues are positive, as
the prior forces the baseline average to zero. The actual noise
average, however, remains undetermined, as in conventional
destriping. 

The sequence of solved baselines does not exactly
follow that of reference baselines,
even if the global offset is subtracted.
This is due to the white noise component in the TOD,
which causes uncertainty
in the determination of the baseline amplitudes.
The combined $1/f$+white noise curve,
binned over the baseline length, is shown Fig.~\ref{fig:todfit}
along with the recovered and reference baselines.

\subsection{Pixelization}
\label{sec:pixelization}

The division into the destriping and binning phases is a
characteristic of the destriping technique. It opens interesting
possibilities for handling signal error, or  gaps in the TOD. By
signal error we mean the error due to temperature variations within
a sky pixel.

Destriping and binning may be carried out at different resolutions.
{\small MADAM}\ makes use of the HEALPix pixelization
\footnote{{\it http://healpix.jpl.nasa.gov}},
where the sky is divided into $12\times nside^2$
equal-area pixels. The baseline amplitudes are first solved
at a {\it destriping resolution}, defined by parameter {\it nside\_cross}.
The solved baselines are then subtracted from the TOD,
and the cleaned TOD is binned into a map at a {\it map resolution},
defined by parameter ${\it nside\_map}$,
which may be smaller or larger than {\it nside\_cross}.
We study the effect of varying {\it nside\_cross}\ in Sect. \ref{sec:resolution}.

{\small MADAM}\  allows to exclude a part of the sky, for instance the galaxy,
defined by an input mask,
in the destriping phase, while including all pixels in the final map.
This feature may be used to reduce the effect of signal error,
which for a large part comes from the foreground-dominated parts of the sky.
Signal error is discussed in Sect. \ref{sec:residualerror}.

In our first paper \citep{Keihanen2005} we considered only
temperature measurements. In that case it was possible to determine
a temperature value for every sky pixel that is hit by any detector
at least once. An additional complication arises when polarization
is involved. To accurately determine the three Stokes parameters for
all pixels, a sufficient coverage in polarization angle is needed in
every pixel. If that is not the case, the Stokes parameters in a
pixel become degenerate.  This may happen for instance, if a pixel
is visited by only one {\sc Planck}\ detector pair. Such pixels may still
be used for determination of the baseline offsets. {\small MADAM}\ does this
by dropping the (nearly) zero eigenvalues and the corresponding
eigenvectors from the pixel matrix in the destriping phase. In the
binning phase the degenerate pixels must be excluded entirely.

\subsection{Gaps}

%
%
\begin{figure}
    \includegraphics[width=9cm]{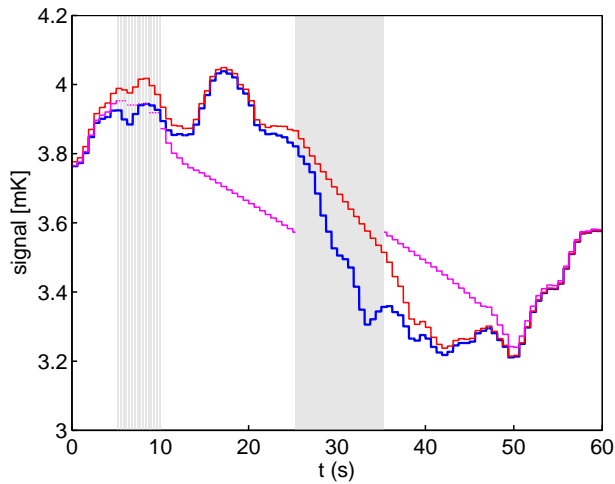}
\caption{
{\bf Effect of gaps.}
We fit 0.625 s (48 samples) baselines to a region where
part of the data is missing. We indicate the gaps by shading.
There is one wider gap (10 s), centered around $t=30$ s, and 16 narrow ones (12 samples).
The thick {\it blue} curve shows the sequence of recovered baselines
in the case all data is there.
The {\it red} curve shows the recovered baselines when the gaps are inserted.
For comparison we show ({\it purple} interrupted curve)
a case where the data is appended
end-to-end and treated as continuous.
}
\label{fig:gap1}
\end{figure}

%
%
\begin{figure}
    \includegraphics[width=9cm]{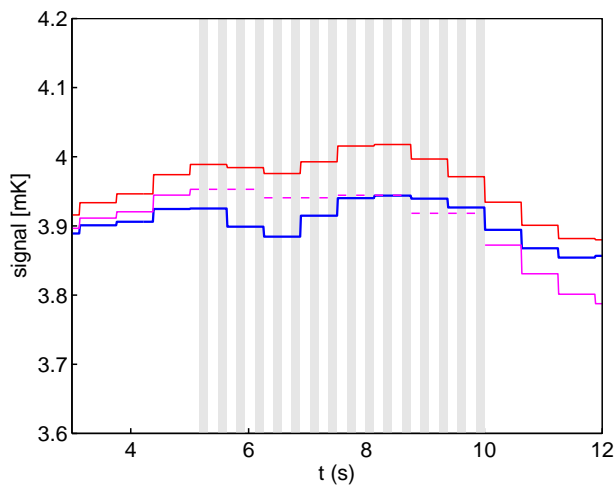}
\caption{
Blow-up of the 3-12 s region of Fig. \ref{fig:gap1}.
}.
\label{fig:gap2}
\end{figure}


The division of noise into a white noise component and a correlated
noise component offers a natural way of handling gaps in TOD. By a
gap we mean a sequence of samples which are missing from the
continuous TOD stream. The data may be corrupted and useless for
analysis, or missing entirely. If the data sections on both sides of
the gap are appended end-to-end, or filled, gaps destroy the
assumption of noise stationarity, which leads to boundary effects
around the gap. A more clever way to handle gaps is to flag the
corrupted samples, and to set the variance of the white  noise
component to infinity for the flagged samples. Remember that we
assumed that the correlated noise component is stationary, but we
did not make the same assumption for the white noise component. 
In practice this done by pointing the flagged samples into a dummy pixel,
which is excluded in matrix operations which involve picking a TOD from a map.
The flagged samples are effectively excluded from map-making, while the
correct statistical properties of the correlated noise component are
still preserved across the gap.

Two fundamentally different cases may be distinguished.
If a gap is longer than the chosen baseline length, some baselines fall completely inside it.
{\small MADAM}\ determines a constrained realization of 
amplitudes also for these baselines,
with the help of the noise prior and the available data on both sides of the gap.
The baselines inside the gap do not contribute directly to the final map, but they have an
effect on the baselines solved on both sides of the gap.
If no noise prior is used, baseline amplitudes become zero inside the gap.

If a gap is shorter than the baseline, some samples on the baseline are missing,
but the baseline amplitude may still be determined based on the remaining samples,
of course with less accuracy.

We demonstrate this in Figs.~\ref{fig:gap1} and  \ref{fig:gap2}. We
have inserted one "large" (10 s) gap and 16 small (12 samples, or 0.15
s) gaps. We show the sequence of baselines in the case all data is
there, and in the case of gaps.

In Fig. \ref{fig:gap1}  we show a one minute chunk of solved 0.625 s
baselines, with all the data present, or with gaps inserted. This is
the same data chunk as in Fig.~\ref{fig:todfit}. For comparison we
also show what happens if the data is appended end-to-end over the
gap. In the latter case, the discontinuity in data leads to a strong
boundary effect, which destroys the fit on both sides of the gap. No
such boundary effect is present when the gap is handled as explained
above.

In Fig.~\ref{fig:gap2} we show a blow-up of the region where we have
inserted gaps shorter than the baseline. The effect is less dramatic there.

\subsection{Split-mode}

Map-making with a short baseline requires a large run-time memory.
One way of reducing the memory requirement is the \emph{split-mode}.
Data is split into small sections, each of which is destriped
independently. Destriping in small pieces tends to leave the lowest
frequency component of each piece poorly determined,  which shows as
relative offsets in different parts of the sky, when the data
sections are combined into one map. We therefore re-destripe the
already cleaned TOD, this time simultaneously, but with a long
baseline (typically 1 hour).
The total
memory requirement in split-mode can be pushed significantly lower
than in standard mode, at the cost of increased CPU time
requirement. Also the map quality suffers slightly as compared to the
standard mode.

\subsection{Computational resources}

The memory requirement of the code is in most cases dominated by
the storage of pointing information. Our simulation data set contains 16
months of data for 4 detectors, at a sampling frequency of 76.8 Hz.
The pointing information for each sample consists of a pixel number
(4 bytes) and factors $\cos(2\psi)$ and $\sin(2\psi)$ (4 bytes
each). The pointing information for the whole 488-day data set takes
$488\times24\times3600\times76.8\times4\time12$ bytes, 145
gigabytes alltogether.

TOD is read into a buffer a small piece at a time, and binned to form the right-hand-side
of the destriping equation. Pointing data instead, must all be kept in memory simultaneously.

When the baseline length exceeds the scanning period, several
samples on one baseline may fall on the same pixel. In such a
situation {\small MADAM}\ combines the samples to reduce the memory
requirement. The compression is lossless, i.e. the compression has
no effect on the output map, just on the resource requirement.
Compression reduces the memory requirement significantly when
using long baselines.

%
%
\begin{table}
\caption[a]{{\bf Computational resources} for various combinations
of input parameters.
}
\begin{center}
\begin{tabular}{rrrrrr}
\hline
\multicolumn{6}{l}{{\bf Varying baseline, noise prior OFF}. Nside=512} \\
\hline
Baseline (s)& Iter & Mem (GB) & Procs & Time (s)& CPUh\\
\hline
3600 &  17  &  7.1 &  32 & 340.6 & 3.03 \\
 600 &  17  & 19.1 &  32 & 287.5 & 2.56 \\
  60 &  17  & 89.1 & 128 & 125.3 & 4.46 \\
  10 &  30  & 89.8 & 128 & 147.2 & 5.23 \\
 2.5 &  87  & 92.3 & 128 & 184.8 & 6.57 \\
1.25 &  131  & 95.5 & 128 & 267.0 & 9.49 \\
\hline
\multicolumn{6}{l}{{\bf Varying baseline, noise prior ON}. Nside=512} \\
\hline
Baseline (s)& Iter & Mem (GB)& Procs & Time (s)& CPUh\\
\hline
  60 &  17  & 89.1 & 128 & 123.6 & 4.39 \\
  10 &  24  & 89.8 & 128 & 138.9 & 4.94 \\
 2.5 &  30  & 92.3 & 128 & 154.7 & 5.50 \\
1.25 &  32  & 95.6 & 128 & 172.0 & 6.12 \\
0.625 & 33  & 102.1 & 128 & 166.1 & 5.91 \\
0.156 & 36  & 142.6 & 256 & 214.4 & 15.24 \\
\hline
\multicolumn{6}{l}{\bf Varying destriping resolution} \\
\multicolumn{6}{l}{Baseline length 0.625 s, noise prior ON.
Nside\_map=512.} \\
\hline
Nside\_cross & Iter & Mem (GB)& Procs & Time (s)& CPUh\\
\hline
 2048 & 35 & 124.7 & 256 & 390.9 & 27.80 \\
 1024 & 35 & 106.6 & 128 & 320.5 & 11.4 \\
 512 & 33 & 102.1 & 128 & 243.7 & 8.67 \\
 128 & 33 & 102.6 & 128 & 236.6 & 8.41 \\
\hline
\multicolumn{6}{l}{{\bf Split-mode}} \\
\multicolumn{6}{l}{Baseline length 0.625 s, noise prior ON, nside=512.} \\
\hline
Split & & Mem (GB)& Procs & Time (s)& CPUh\\
\hline
2 & &  55.1  & 128 &  267.1 & 9.50 \\
4 & &  29.9  & 128 &  348.1 & 12.38 \\
 8 & &  16.7  &  32 & 1288.9 & 11.46 \\
 61 & &   5.7  &  32 & 2059.8 & 18.31 \\
 \hline
\end{tabular}
\end{center}
\label{tab:resources}
\end{table}

All our simulations were run on the Cray XT4/XT5 "Louhi" system of CSC (IT Center for Science), 
Finland. We used from 32 to128 processors. 
The number of processors was selected according to the memory requirement.

In Table \ref{tab:resources} we list the computational resources for various runs.
In all the listed cases, we built intensity and polarization maps of
the full data set (488 days, 4 detectors at sampling frequency of 76.8 Hz).
We vary the baseline length and destriping resolution.
In the first column we show the parameter we have varied, that is the baseline length,
resolution in the destriping phase,
or the number of data chunks in split-mode.
The other columns show the number of conjugate-gradient iteration steps,
total memory usage, number of processors,  and CPU time usage as wall-clock time and as
CPU hours.

The memory requirement depends strongly on the chosen baseline
length at long baselines (above 60 s) because the effectiveness of
data compression strongly depends on how many samples on a baseline
fall on the same pixel. 

The split-mode offers a trade-off between memory and CPU time usage.
When data is destriped in small chunks, memory usage can be pushed
very low, but at the same time the required CPU time increases
drastically.

%
%

\section{Simulations}

%
%
\begin{figure}
     \includegraphics[width=9cm]{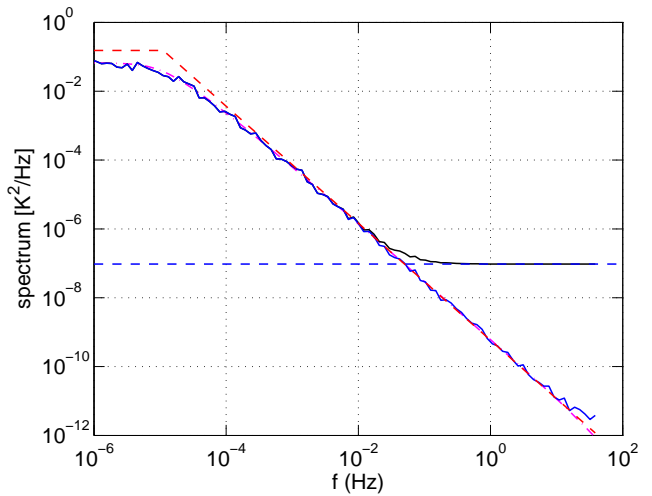}
\caption{
{\bf Noise spectrum.}
The {\it dashed red} line shows the analytical $1/f$ noise model used when
constructing the noise prior.
The {\it solid blue} line gives the actual spectrum of the input noise,
computed as an average over 32 noise realizations.
The {\it dash-dotted purple } line presents the numerically predicted SDE spectrum.
We show also the white noise spectrum ({\it dashed blue}) and the
sum of the white and $1/f$ noise ({\it solid black}).
}
\label{fig:sdespec}
\end{figure}
%

%
%
\begin{figure}
     \includegraphics[width=5.5cm,angle=90]{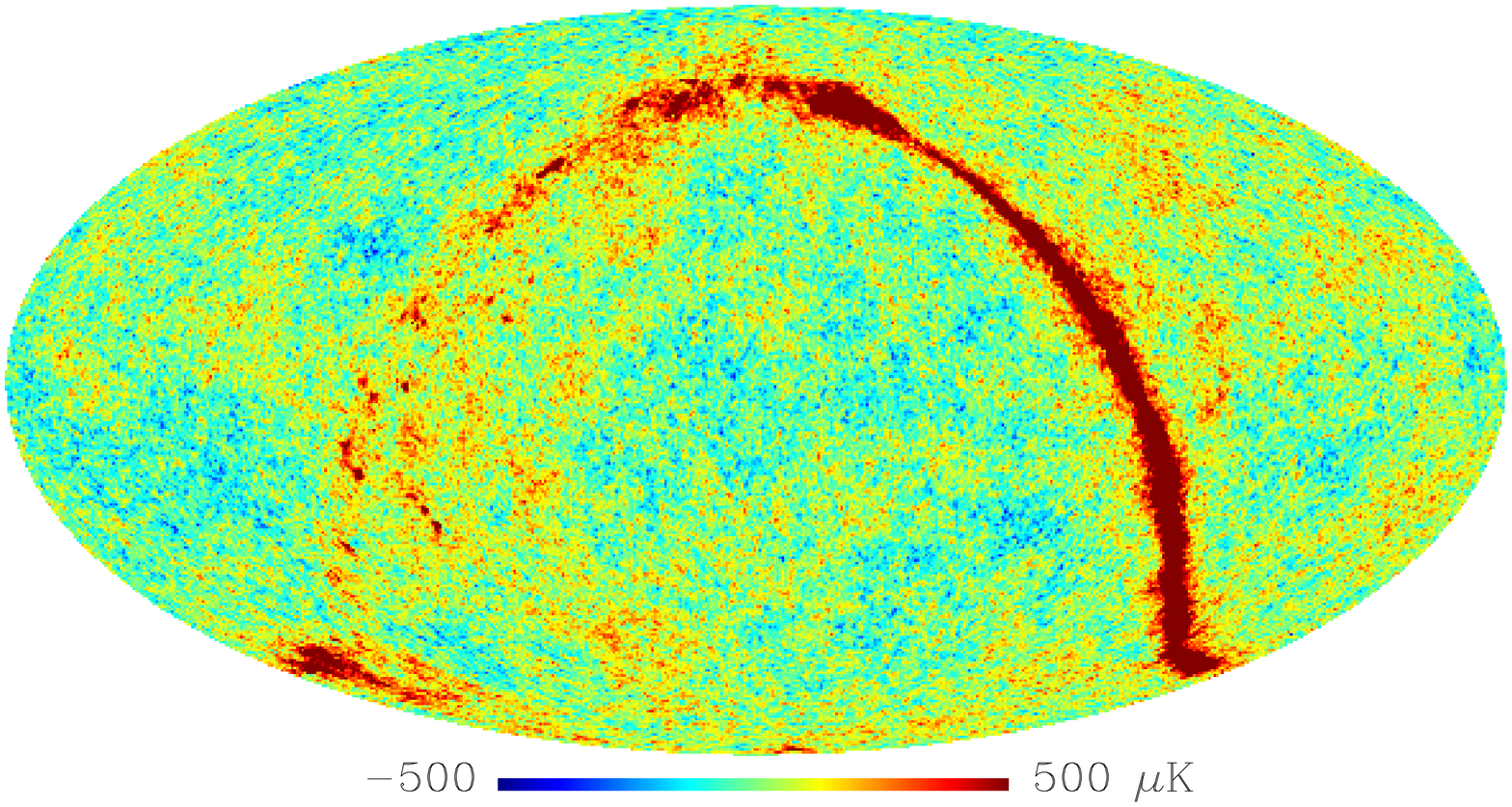}
     \includegraphics[width=5.5cm,angle=90]{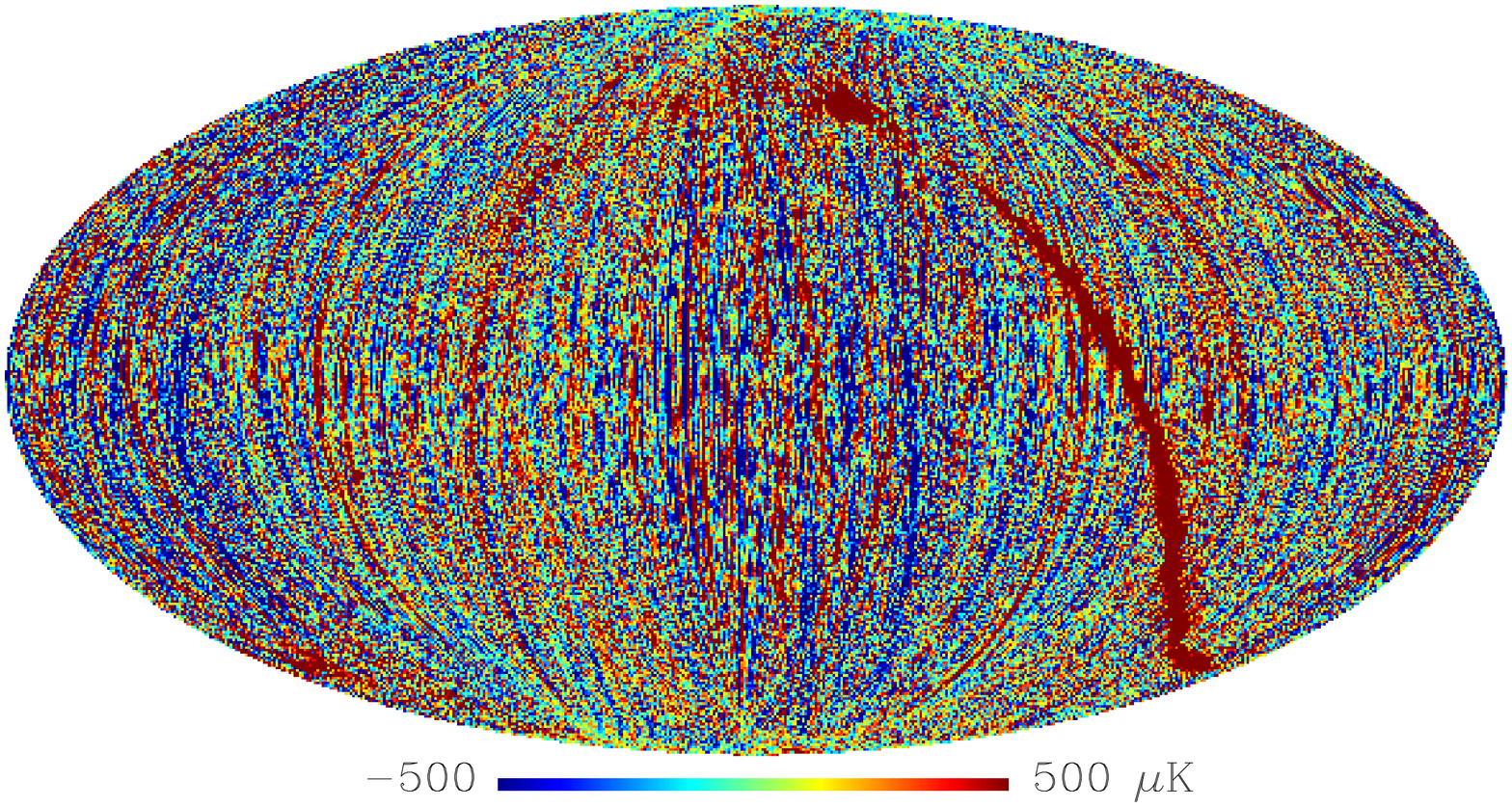}
     \includegraphics[width=5.5cm,angle=90]{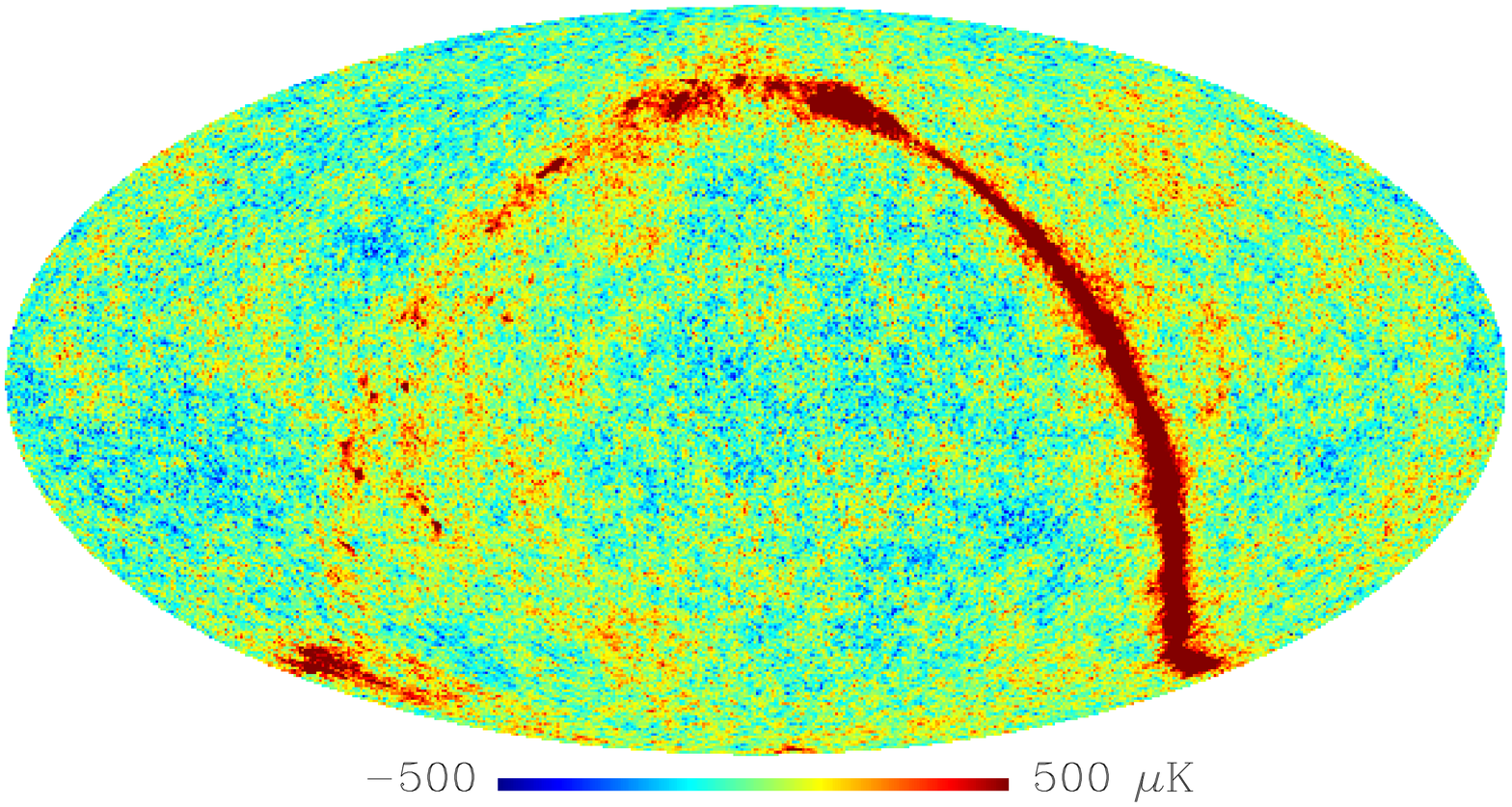}
\caption{
{\bf Full-sky maps.}
From top to bottom:
Binned noiseless map (total intensity),
map directly binned from TOD,
and destriped map (baseline length 0.625 s).
The binned map is contaminated by stripes due to $1/f$ noise.}
\label{fig:maps}
\end{figure}
%

We created a set of simulated time-ordered data,
mimicking the data from
the Planck LFI (Low-Frequency Instrument) 70 GHz channel.
The TOD was produced using {\sc Planck} LevelS simulation software \citep{Reinecke2006}.
We created data for 16 months for four detectors,
corresponding to the LFI horns 19 and 22.

We ran {\small MADAM} on this simulation data with different parameter
settings. Results are discussed in the following sections.

For each of the four detectors, we created four TOD streams of 16 months duration:
a CMB signal, a foreground signal, $1/f$ noise, and white noise.
Each TOD consisted of
$488\times24\times3600\times76.8=3.24\cdot10^9$ samples.
Storing the TOD components separately allowed us to do noise-only and signal-only
simulations.

Pointing information was stored  in the form of satellite pointing,
sampled at 1 s intervals. {\small MADAM}\ constructed detector pointings,
sampled at the sampling frequency 76.8 Hz, with the help of
focal plane parameters.

The same data set was used by \cite{HX2009}, with the exception that
in this work we used 16 months of data (instead of 12). We have also
included a foreground component.

\subsection{Scanning}

The scanning strategy imitates that of the actual {\sc Planck} spacecraft \citep{Dupac2005}.
The spin axis rotated around
the anti-Sun direction once every six months, in a circle of radius
$7.5\degr$.
When projected onto the sky, this creates a cycloidal path.
The spin axis was repointed at fixed intervals of one hour. Between
repointings, the spin axis nutated with a mean amplitude $1.6'$.

The detectors scanned the sky almost along great circles. The
satellite rotated around the spin axis with a mean rate of
$f_\mathrm{sp}=1/60$ Hz (one rotation per minute). The detectors were
pointed $\theta_\mathrm{det}=87.77\degr$ away from the spin axis.
Small variations (rms 0.1$\degr$/s) were added to the spin rate.
Because of the nutation and the spin rate variations, adjacent
scanning circles did not fall exactly on top of each other.

The two detectors belonging to the same horn shared the same
pointing ($\theta,\phi$) on the sky, but the polarization angles
were different by 90.0$\degr$. The two horns followed the same path,
but one horn trailed the other by 3.1$\degr$. The polarization
angles of the two horns were roughly at 45$\degr$ angles, so that
the polarization measurements of the two horns complemented each
other. For a more detailed description of the scanning strategy see
\cite{HX2009}.

During the 16 month observation time the four detectors covered each pixel
on the sky at resolution {\it nside\_map}=512 in multiple polarization directions.
Therefore we were able to determine the three Stokes components in every pixel.

\subsection{Signal}

Our simulation data set contained signal from CMB and from
foregrounds. 
No dipole signal was included.
We used the CAMB\footnote{http://camb.info} code to
produce a theoretical CMB angular power spectrum with cosmological
parameter values $\Omega_0=1, \Omega_\Lambda=0.7, \omega_m=0.147,
\omega_b=0.022, \tau=0.1$. We then created a realization of
coefficients $a^\mathrm{T}_{lm}$ and $a^\mathrm{E}_{lm}$. No B mode polarization was
included. We constructed a TOD with sampling frequency
$f_\mathrm{s}=76.8$ Hz, smoothing with a symmetric Gaussian beam with
FWHM $= 12.68'$ \citep{Wandelt2001,Challinor2000}.

The foreground signal included Galactic emission from thermal and
spinning dust, synchrotron radiation, and free-free scattering. On
top of the galactic signal we added a Sunyaev-Zeldovich signal, and
weak point sources. We constructed an $nside=2048$ sky map including
these components, using the Planck sky model, PSM, version 1.6.3
\footnote{
http://www.apc.univ-paris7.fr/APC\_CS/Recherche/Adamis/PSM/psky-en.php}.
The map was smoothed with a symmetric gaussian beam.
We then constructed the foreground TOD
by picking values from the input map according to the scanning
pattern.

\subsection{Noise}

We generated $1/f$\ noise by the SDE (Stochastic Differential Equation) algorithm,
which builds the noise stream as  a linear combination of a number
of low-pass filtered white noise streams
\citep{Reinecke2006}.
We used the following input parameters:
knee frequency of $f_\mathrm{kn}=0.05$ Hz, sampling frequency
$f_\mathrm{s}=76.8$ Hz, slope $\alpha=-1.7$, and minimum frequency
$f_\mathrm{min}=1.15\times10^{-5}$ Hz.
On top of the $1/f$\ noise we added Gaussian white noise with a standard deviation of
$\sigma=2.7$ mK and zero mean.

We have chosen a high knee frequency in order to show clearly the
effect of correlated noise in the output maps. The purpose of these
simulations is not to demonstrate the actual performance of the
{\sc Planck} experiment, but to give a quantitative picture of how
changing the various parameters which control the {\small MADAM} algorithm
affects the quality of output map and the required computational
resources.

We show the noise spectrum in Fig.~\ref{fig:sdespec} together with
the analytical model used to construct the noise prior. We computed
the noise spectrum through the Fourier technique from noise streams
of 366 hours, averaging over 32 independent noise realizations, at
100 distinct frequencies. The analytical model describes the noise
well at high frequencies, but differs somewhat at low frequencies.
The analytical model is given by 
\begin{eqnarray}
   P(f) &=& \frac{\sigma^2}{f_s}\left( \frac{f}{f_\mathrm{kn}}\right)^\alpha \qquad (f >f_\mathrm{min}) \\
   P(f) &=& \frac{\sigma^2}{f_s}\left( \frac{f_\mathrm{min}}{f_\mathrm{kn}}\right)^\alpha \quad\, (f <f_\mathrm{min}) \nonumber
\end{eqnarray}
It is possible to theoretically predict the spectrum produced by the SDE algorithm.
The prediction is plotted in the same figure,
and agrees well with the actual spectrum.

%
%

\section{Residual error}
\label{sec:residualerror}

%
%
\begin{figure}
     \includegraphics[width=5.5cm,angle=90]{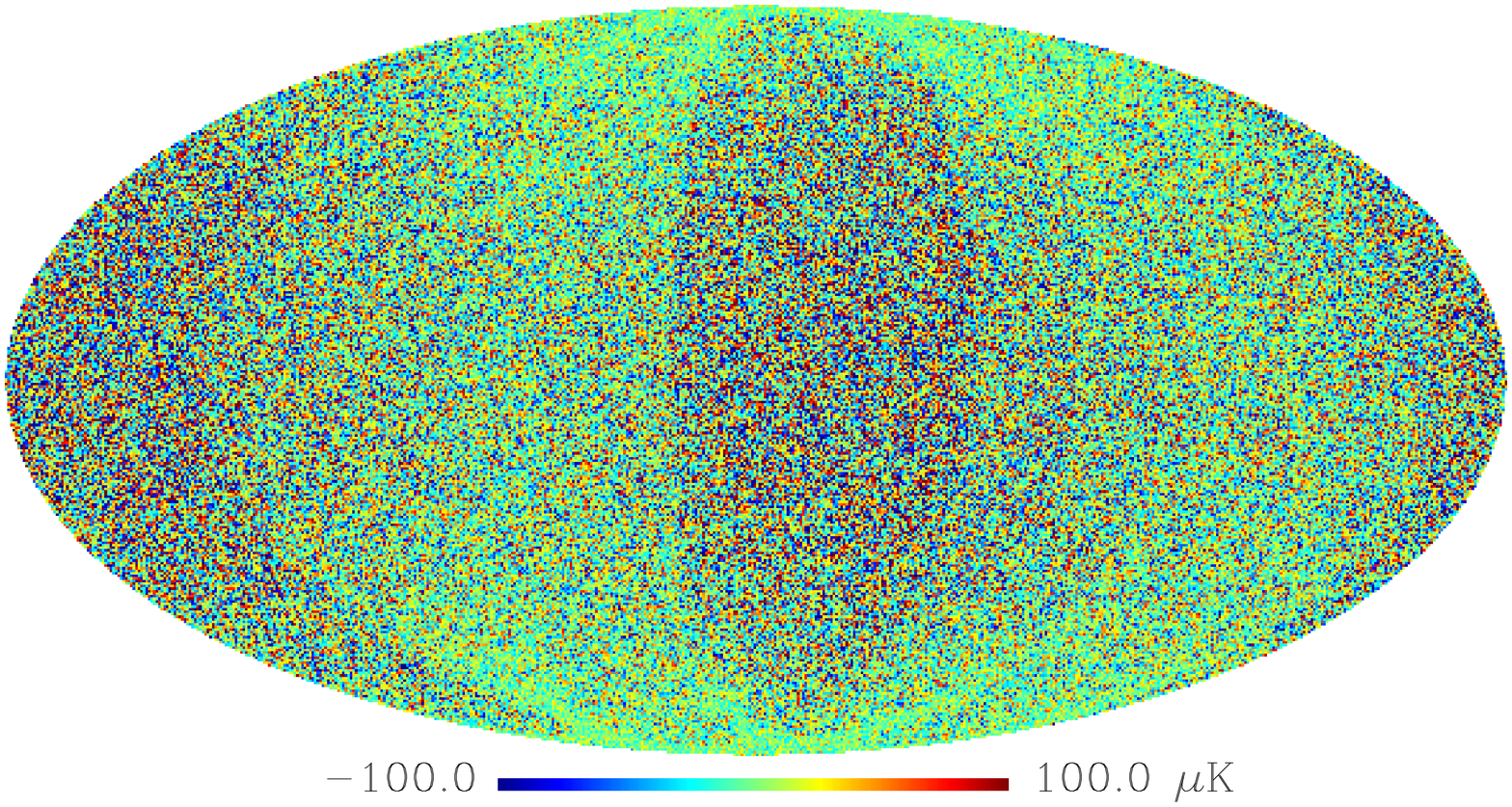}
     \includegraphics[width=5.5cm,angle=90]{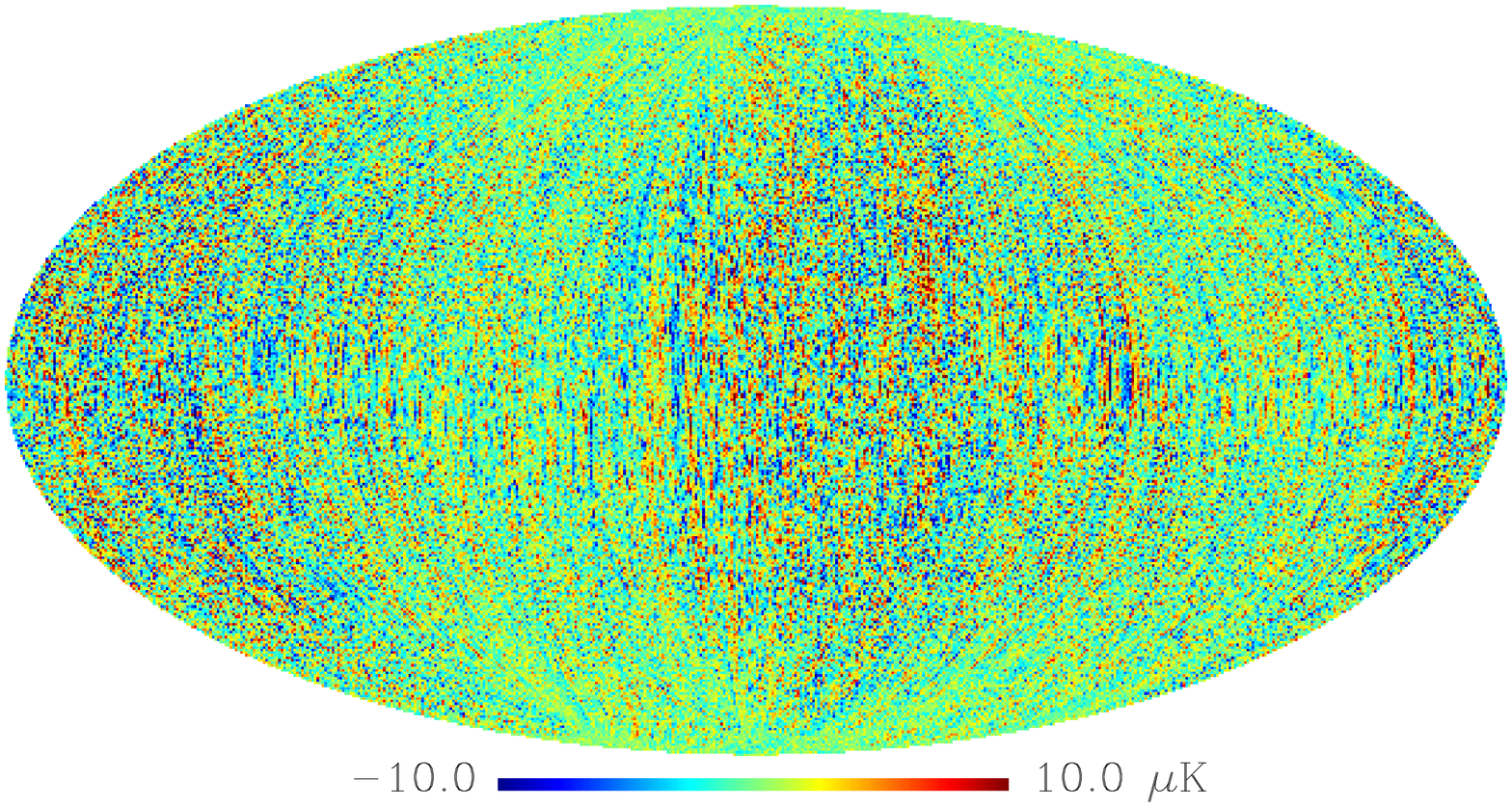}
\caption{
{\bf Residual noise.}
Upper map shows the residual noise
map.  The bottom map shows the correlated residual, which we obtain
by subtracting the binned white noise from the residual noise map.
Note the difference in scale.} \label{fig:residualmaps}
\end{figure}


%
%
\begin{figure}
\includegraphics[width=9cm]{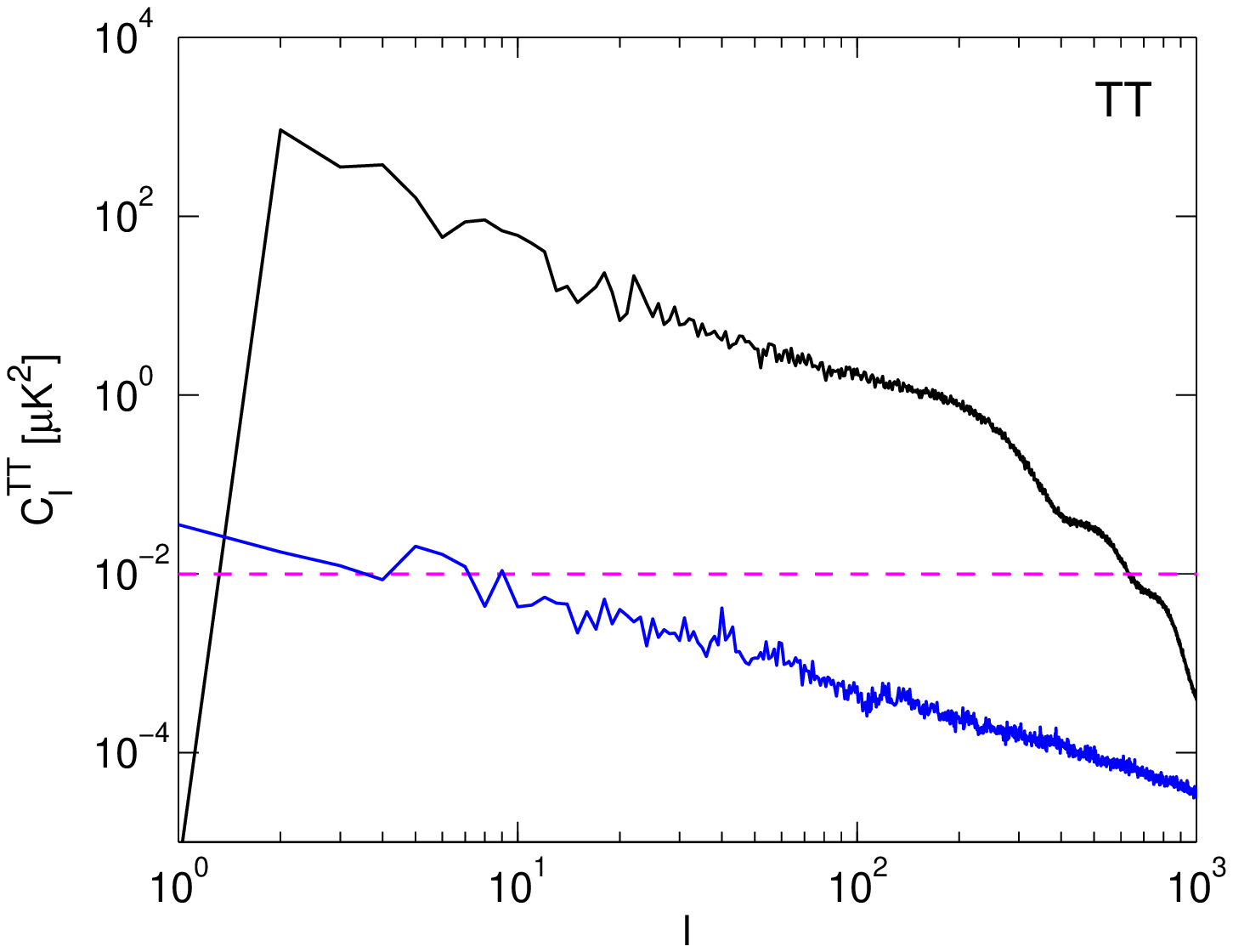}
\includegraphics[width=9cm]{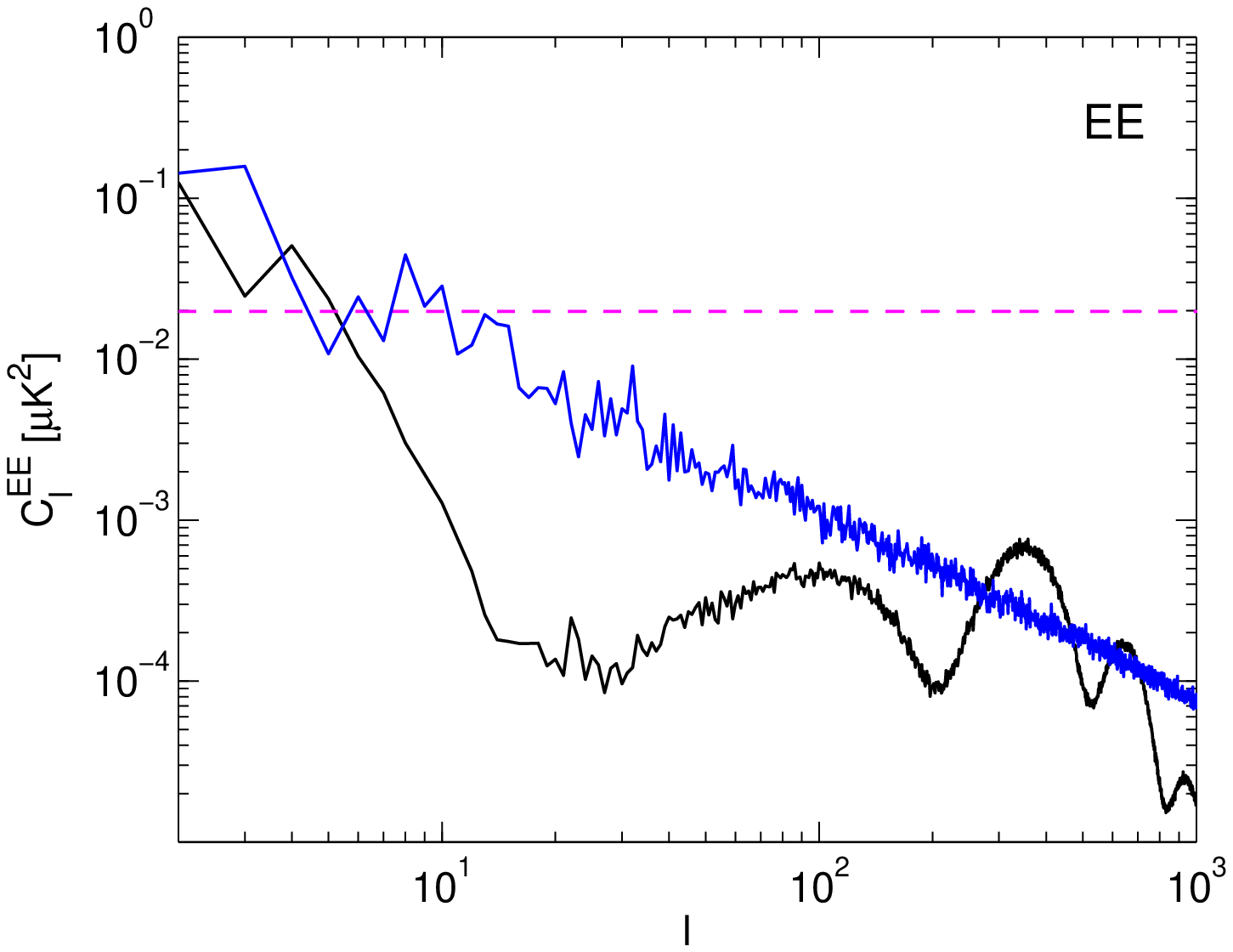}
\caption{
 {\bf Angular power spectrum of signal and residual noise}.
We show TT and EE spectra of the correlated residual noise (CRN) map
({\it blue} curve), together with the spectra of the binned
noiseless CMB map {\it black}. Also shown is the white noise level
({\it purple dashed} line). The baseline length was 0.625 s.}
\label{fig:clspec}
\end{figure}


To assess the quality of the destriped map, we study first the {\it
residual error map}, which we compute as the difference between the
destriped map and the {\it binned noiseless map}. The binned
noiseless map is obtained by binning the simulated signal TOD into a
map as by Eq. (\ref{binmap}).

The top panel of Fig. \ref{fig:maps} shows the binned noiseless map.
We show also the map binned from noise-contaminated TOD, and the destriped map.
The map binned from noisy data is contaminated by stripes due to $1/f$ noise.
The difference between the bottom and top panels is the residual error map.

The residual error map can further be divided into two statistically
independent components: the {\it signal error map} and the {\it
residual noise map}. The signal error map is obtained by destriping
the signal-only TOD and subtracting the binned noiseless map from
it. The residual noise map is obtained by destriping the noise-only
TOD. Because destriping is a linear process, the total residual
error is a sum of signal error and residual noise.

Signal error arises from temperature variations within a pixel, which the code interprets
as noise and which lead to non-vanishing
baseline amplitude estimates even in absence of noise.
The signal error has the undesired effect of spreading sharp
features in the signal map along the direction of the scanning ring.

Residual noise is strongly dominated by white noise. Though it is
the dominant component, it is rather simple to handle at later data
processing levels, for instance power spectrum estimation. We are
more interested  how much {\it correlated} noise there is left in
our output maps. We therefore subtract the  {\it binned white noise
map} from the residual noise map, and study the remaining {\it
correlated residual noise} (CRN) map separately.

The residual noise map, as well as its correlated component, are
depicted in Fig.~\ref{fig:residualmaps}.

It can be shown that the CRN map and the binned white noise map are
statistically independent, if the final map is binned to a
resolution equal to or lower than the destriping resolution. The
proof was presented by \cite{HX2009} for the case of equal
resolutions. Here we show that the maps are independent also if the
destriping resolution is higher than the map resolution.

The binned white noise map is proportional to the quantity
\begin{equation}
    \vec m_\mathrm{b}\propto \mathbf{P}_\mathrm{m}^T\mathbf{C}_\mathrm{n}^{-1}\vec w
\end{equation}
where $\vec w$ is the white noise TOD. Since the binned white
noise is obviously uncorrelated with $1/f$ noise and signal, it is
enough to study its correlation with the CRN map one obtains
destriping a white noise TOD alone. This is proportional to the
quantity
\begin{equation}
     \vec m_c\propto  \mathbf{Z}\vec w
        = (\mathbf{I}-\mathbf{P}_\mathrm{c}(\mathbf{P}_\mathrm{c}^T\mathbf{C}_\mathrm{n}^{-1}\mathbf{P}_\mathrm{c})^{-1}
                                \mathbf{P}_\mathrm{c}^T\mathbf{C}_\mathrm{n}^{-1} )\vec w
\end{equation}
The correlation between the two maps is proportional to
\begin{equation}
  \langle \vec m_\mathrm{c}  \vec m_b\rangle
     \propto \mathbf{Z} \langle\vec w\vec w^T\rangle \mathbf{C}_\mathrm{n}\mathbf{P}_\mathrm{m}
       = \mathbf{Z}\mathbf{P}_\mathrm{m} .
\end{equation}
From the definition of $\mathbf{Z}$ one sees that $\mathbf{Z}\mathbf{P}_\mathrm{c}=0$.
If now the map resolution is lower than or equal to the destriping resolution,
one can write $\mathbf{P}_\mathrm{m}=\mathbf{P}_\mathrm{c}\vec{S}$,
where the effect of matrix $\vec S$ is to downgrade
matrix $\mathbf{P}_\mathrm{c}$ to a lower resolution by summing columns.
One then readily sees that the correlation vanishes.

If the map is binned to a resolution {\it higher} than the
destriping resolution, we have the opposite relation
$\mathbf{P}_\mathrm{c}=\mathbf{P}_\mathrm{m}\vec{S}$. 
In that case, the correlation between the CRN
map and the binned white noise map does not vanish. This has some
interesting consequences, which are discussed in Sect.
\ref{sec:resolution}.

Residual noise does not share the symmetry of the original noise.
Residual noise is not stationary in the time domain, like the
original $1/f$ noise. Note also that no noise component is
statistically isotropic in the map domain.
 A complete description of the residual noise requires the calculation
 of the noise covariance matrix (NCVM), which is outside the scope of this paper.
 Computation of the NCVM for low-resolution maps is discussed by \cite{Keskitalo2009}.

 In this paper we calculate some simple figures of merit,
 which describe different aspects of the residual noise.
In cases where the binned white noise map and the CRN map are
independent, the variance  of the CRN map is a useful number.
 It has the benefit of describing the residual noise level as one number,
 which can be plotted as a function of various parameters.
 We calculate the variance of the CRN map for the three Stokes maps, I,Q,U, separately.

Variance is an additive number.
When the residual error map can be divided
into independent components, we compute and plot the variance of each individual
component separately. The total variance of the residual error is obtained as the sum
of variances of the individual components.

When destriping is performed at a resolution lower than the
map resolution, the CRN map and binned white noise are correlated.
In this situation also the covariance between the two maps is an
important quantity. The covariance is expressed in the same
units as the variance, which makes it easy to compare the two. This
is the main reason why we have chosen to use the variance as the
main figure of merit, instead of rms. 
The covariance is computed by averaging over
all sky pixels the product of temperature values of the two maps in
question.

We calculate also the spectrum of the residual noise
in time domain, as a square of the
 Fourier transform of the noise stream. Neither this is a complete description of the
 residual noise, because the residual noise is not stationary (not diagonal in Fourier domain).

 Finally, we calculate the angular power spectrum of the residual noise map.
 Again, because the residual noise is not statistically isotropic,
 the angular power spectrum does not fully describe the noise,
 but gives an idea at which scales the noise is distributed in the map domain.
 It also gives the {\it noise bias} which should be subtracted from the spectrum
 of the destriped map to get an unbiased estimate of the power spectrum of the
 underlying signal map.

In Fig.~\ref{fig:clspec} we show first the spectrum of the CRN map
for 0.625 s baseline length, together with the spectrum of the
binned noiseless map. We show also the white noise level.

%
%

\section{Effect of baseline length}
\label{sec:baseline}

\subsection{Map domain}

In this section we study the effect of baseline length on the residual
noise. We destripe and bin the final map at a fixed resolution
{\it nside\_cross}={\it nside\_map}=512. The destriped map consists of 3 million
pixels ($12\times512\times512=3145728$) for each of the three Stokes
components.

Since we destripe and bin the map at the same resolution,
the binned white noise map and
the correlated residual noise (CRN) map are statistically independent.

%
%
\begin{figure}
  \includegraphics[width=9cm]{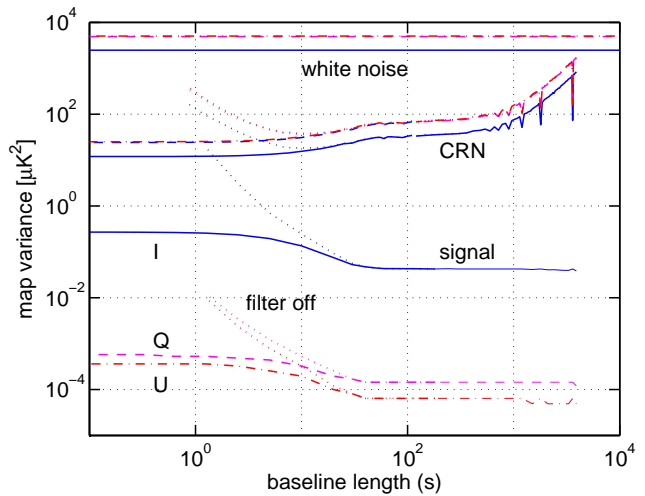}
\caption{
 {\bf Effect of baseline length on residual error.}
 The
residual error map is divided into three components: binned white
noise (above), correlated residual noise (middle), and signal error
(below). We show the variance of each residual component map, as a
function of baseline length. Binned white noise is independent of
baseline length, and its variance is shown by a straight line. We
show the three Stokes parameters I,Q,U ({\it blue solid}, 
{\it purple dashed}, and {\it red dash-dotted} line, respectively). The
{\it dotted} lines show the effect of turning the noise prior off.
At long baselines results with and without noise prior coincide. }
\label{fig:baseline_all}
\end{figure}

%
%
\begin{figure}
\includegraphics[width=9cm]{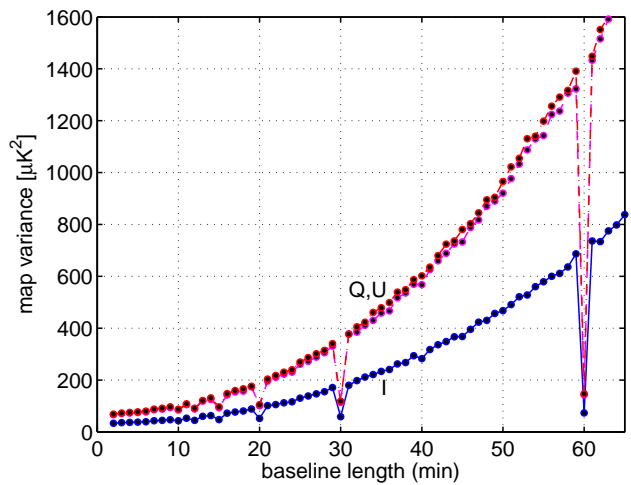}
\caption{
{\bf Long baselines.}
Variance of the CRN map as a function of baseline length,
for long baselines.
This is a blow-up of the long-baseline part of the CRN curves
in Fig. \ref{fig:baseline_all}.
We have computed the residual noise at 1 minute intervals.
Baseline lengths that go evenly into the repointing period of 1 hour,
stand out as local minima.
}
\label{fig:baseline_long}
\end{figure}

%
%
\begin{figure}
\includegraphics[width=9cm]{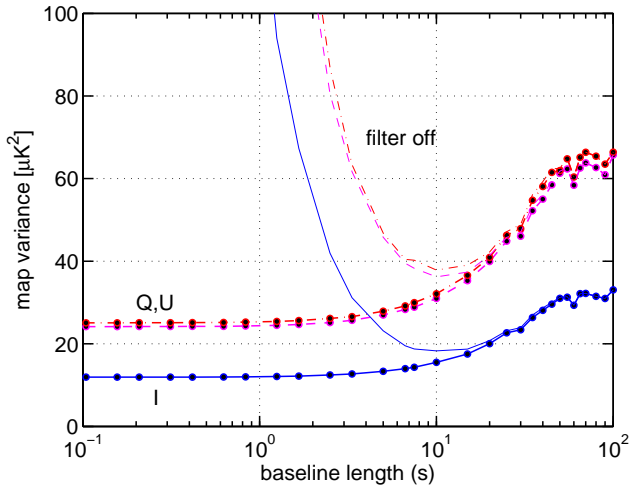}
\caption{
{\bf Short baselines.}
 We show the CRN variance as a
function of baseline legth, for short baselines. This is a blow-up
of the short-baseline part of the CRN curves in Fig.
\ref{fig:baseline_all}. We show again the three Stokes parameters
with different line types. The upper, thin lines show the effect of
turning off the noise filter.}
 \label{fig:baseline_short}
\end{figure}


%
%
\begin{figure}
\includegraphics[width=9cm]{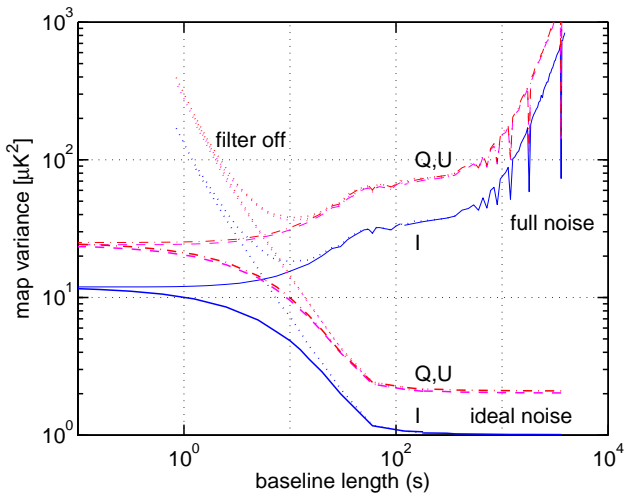}
\caption{
{\bf Ideal noise.}
We show the CRN variance as a function of baseline length for real noise
(same as in Fig. \ref{fig:baseline_all}) and for idealized noise,
composed of reference baselines.
}
\label{fig:baseline_ref}
\end{figure}


In Fig. \ref{fig:baseline_all} we plot the variance of various
components of the residual error map, as a function of baseline
length. We show the variance of the signal error map, CRN map, and
binned white noise map, for the three Stokes parameters (I,Q,U). We
show results both with and without noise prior. The total variance
is the sum of the three components.

We discuss in the following the dependence of the results on
baseline length. Some features are readily visible.

The variance of the binned white noise is, obviously,
independent of baseline length.

As a general trend, correlated residual noise decreases with decreasing baseline length,
as the noise becomes better modelled by the selected baselines.
Baseline lengths which are an integer fraction of the repointing period (1 hour),
give a local minimum.
Below 1 second the results converge.
In the remainder of this paper we often use baseline length 0.625 s (48 samples)
as an example of a very short baseline.
At this baseline length, results have already converged with respect to baseline length,
but requirements for computational resources are still moderate.

As opposite to residual noise, signal error increases with decreasing baseline length,
but remains always below the noise level.
Signal error becomes more important if destriping is performed at a low resolution.
We discuss this in Sect. \ref{sec:resolution}.

The noise prior has a negligible effect on results above 1 minute (
$\gg 1/f_\mathrm{kn}$) baseline length. At short baselines  ($< 1/f_\mathrm{kn}$),
on the contrary, the noise prior becomes important. When the noise
prior is turned off, the required CPU time increases steeply with
decreasing baseline length, as the algorithm requires a large number
of iteration steps to converge. Therefore we have not computed
results without the noise prior at the very shortest baseline
lengths.

The variance of residual noise for Q and U is above that of I,
roughly by a factor of 2. This reflects the fact that Q and U
contribute to the observed signal weighted by factors $\sin(2\psi)$,
$\cos(2\psi)$ (Eq.~\ref{poltod}) relative to I.

The signal error in Q and U is well below that of the temperature
component. This follows directly from the fact that the polarization
signal is weaker than the intensity signal.

To show the behaviour of the CRN curve more clearly, we plot it 
separately for long and short baselines
in Figs.~\ref{fig:baseline_long} and \ref{fig:baseline_short}.

Baseline lengths which are an integer fraction of the repointing
period (here 1 hour), give a clear and sharp local minimum. The minimum
is deepest at 1 hour baseline length, but strong dips can also be
seen at baseline lengths 30 min, 20 min, 15 min, 12 min and 10 min.

The dip structure is obviously related to the scanning strategy. 
Why the use of baselines that fit into destriping periods is better 
than the use of ones that do not, can be understood when thinking of
an ideal scanning, where scanning
rings fall exactly on top of each other. The noise stream can be
thought as composed of two components: an "offset" component
consisting of 1 minute reference baselines, defined as averages of
the noise stream over 1 min periods, and a second component which
contains the remaining high-frequency noise. The offset component
dominates the noise. When the noise stream is averaged over 60
adjacent scanning circles, the offset component produces a residual
offset, equal to the average of 60 adjacent reference baselines.
When the data is destriped with a one-hour baseline, destriping
effectively removes the residual offset. If the baseline length
differs from one hour, a time offset develops between baselines and
repointing periods, with the result that noise offsets do not cancel
out.

In this work we have assumed a scanning strategy, where
the repointing period is constant at 1 hour,
as was originally planned for Planck,
and study only constant baseline lengths.
In reality, Planck is likely to use a scanning pattern
where the repointing period varies.
We have not run simulations with a variable pointing period,
but based on the results presented here,
we expect that for optimal results the baseline length should
follow the repointing period,
if a long baseline length is required.
A varying baseline length has therefore
been implemented in {\small MADAM}
for the use of the {\sc Planck} experiment.

%
%
\begin{figure}
\includegraphics[width=9cm]{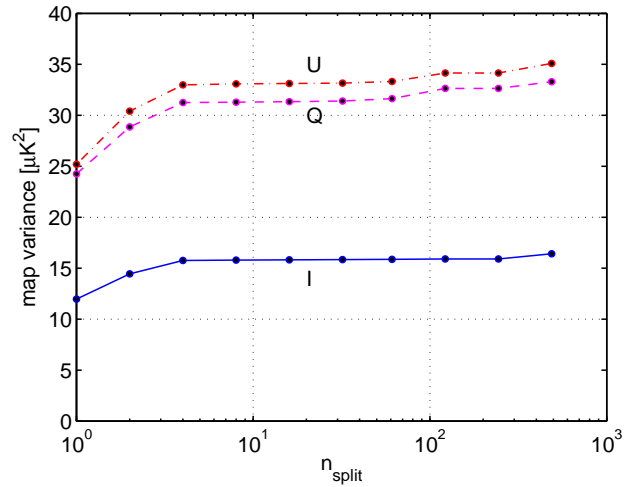}
\caption{
{\bf Split-mode.}
Residual correlated noise variance for
split-mode, as a function of the splitting factor. In the first
phase data was destriped in chunks of length 488 days/$n_{split}$.
The line types correspond
again to three Stokes parameters: I ({\it solid}), Q ({\it dashed})
and U ({\it dash-dotted}).} \label{fig:splitmode}
\end{figure}


When no noise prior is used, the CRN variance as a function of baseline length
has a global minimum, as a result of two
competing effects. When moving towards shorter baselines, noise
becomes better modelled. On the other hand, the shorter the
baselines, the more there are unknowns (baseline amplitudes) to be
solved, while the amount of data remains fixed.
With our data set the optimal baseline length was 10 s.
\cite{HX2009} split
the CRN map further into {residual $1/f$\ noise} and {white noise
baselines} to study these competing effects more deeply. The optimal
baseline length depends on the knee frequency of the $1/f$\
component. 

With the noise prior, results continue to improve at least until
baseline lengths of 0.1 s, which is the limit where we could bring
our computations. The noise prior has the effect of restricting the
baseline solution in such a way that the effective number of unknows
does not follow the number of baselines.

In order to study the importance of the high-frequency part of the
noise spectrum, which is not well modelled by baselines, we made a
simulation where we removed the part of noise not modelled by the
baseline approximation. For each baseline length, we replaced the $1/f$\ noise
component by a sequence of baselines of the same length as the
baseline length used in destriping. 
The results
are shown in Fig~\ref{fig:baseline_ref}, together with results
obtained with realistic noise. The difference between the two sets
of curves is the contribution of noise not modelled by baselines. We
see that this component becomes very important at long baselines.

In Fig~\ref{fig:splitmode} we show results obtained by applying the
split-mode. Data was divided into chunks which were first destriped
separately with a 0.625 s baseline.
In the second phase we combined the chunks,
and re-destriped the data with 1-hour baselines. 
The rightmost points correspond to a case where the data was
destriped in 1-day chunks.
The leftmost point corresponds to the standard case, where the
whole 488-day data set is destriped once. 
 The CRN variance 
in split-mode is
higher than in the standard mode. This reflects the fact
that the split-mode does not exploit all information in the data.
The benefit of the split-mode is that it requires less memory than 
the standard mode, as can
be seen from Table \ref{tab:resources}.

\subsection{Time domain}
\label{sec:base_timedomain}

%
%
\begin{figure}
\includegraphics[width=9cm]{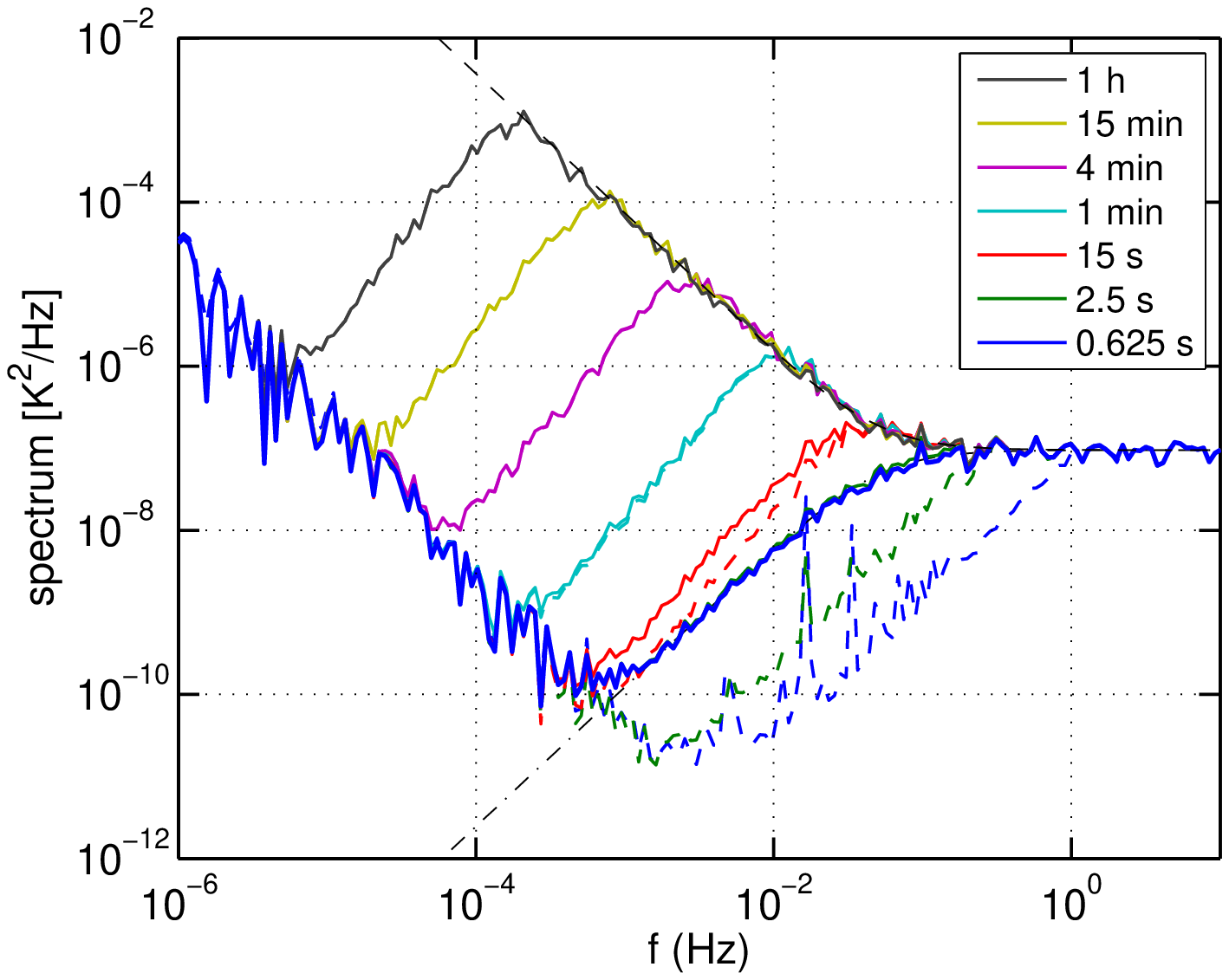}
\caption{
{\bf Effect of baseline length on residual noise spectrum.}
We subtract the solved baselines from the noise TOD (white noise and $1/f$ noise)
and plot the spectrum of the residual.
We show results for 7 baseline lengths, from above:
1 h ({\it black}), 15 min ({\it yellow},
4 min ({\it purple}), 1min ({\it cyan}), 15 s ({\it red}), 2.5 s ({\it green}) and 0.625 s ({\it blue}).
{\it Solid} and {\it dashed} lines show results with and without noise prior,
respectively.
With 1 min baselines and longer, the dashed and solid curves (noise prior on/off)
are on top of each other. The 0.625 s case with noise prior, which of the studied cases
gives the
 lowest residual noise variance in map domain,
is shown by a thick linetype.
Also shown are the original noise spectrum ({\it black dashed})
and the analytical approximation given by Eq. (\ref{analytical}) ({\it black dash-dotted}).
}
\label{fig:todspec1}
\end{figure}


%
%
\begin{figure*}
\hbox{
\includegraphics[width=9cm]{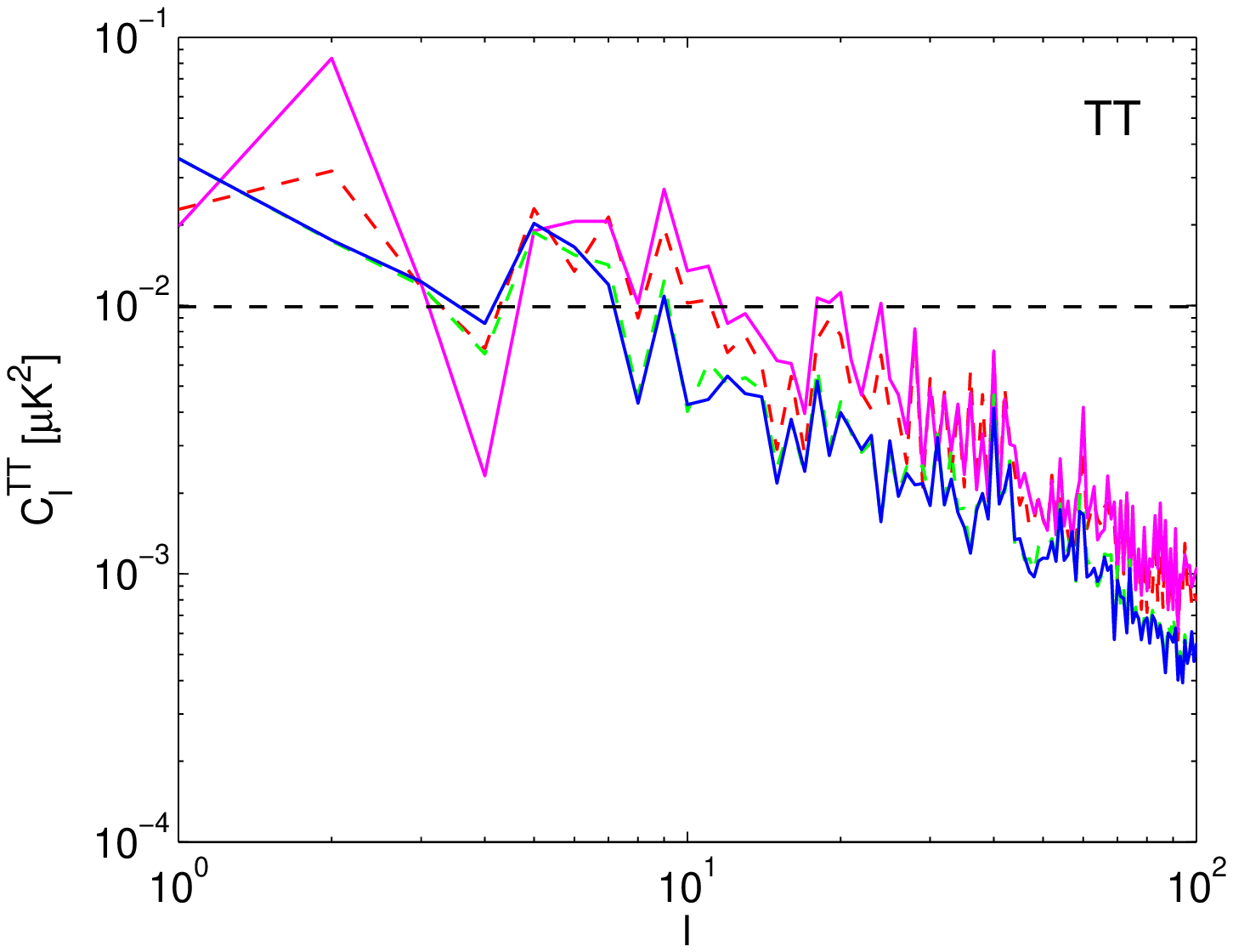}
\includegraphics[width=9cm]{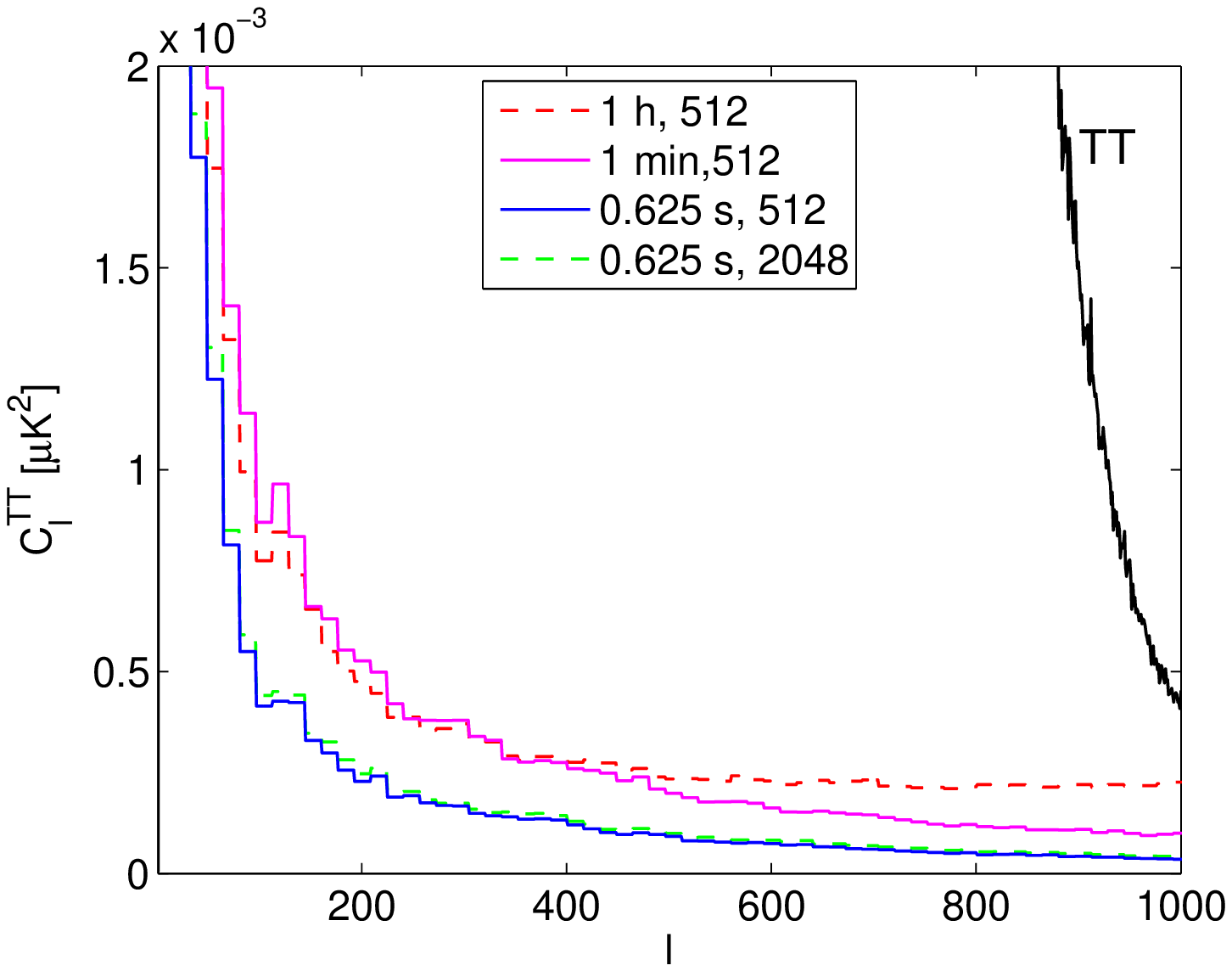}}
\hbox{
\includegraphics[width=9cm]{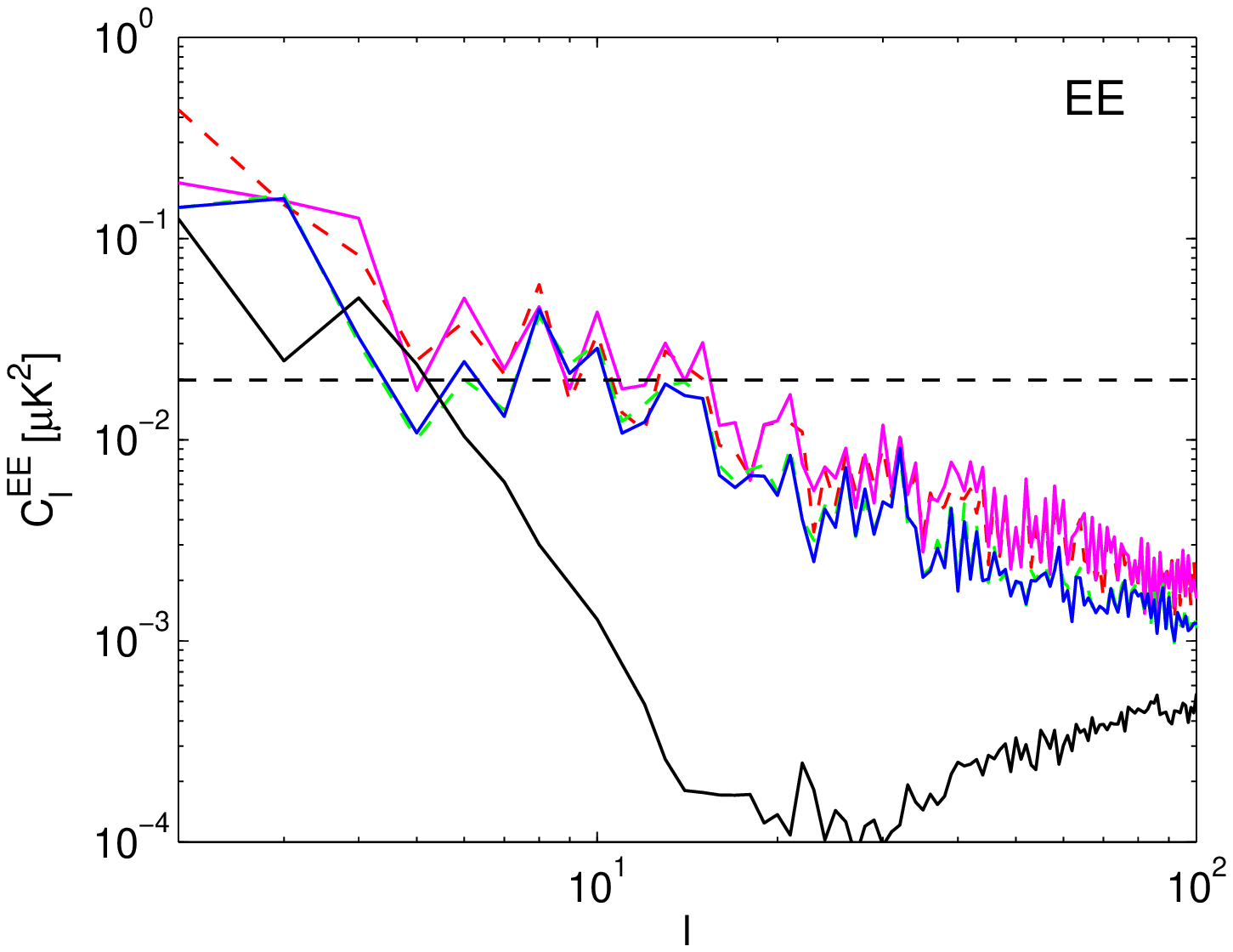}
\includegraphics[width=9cm]{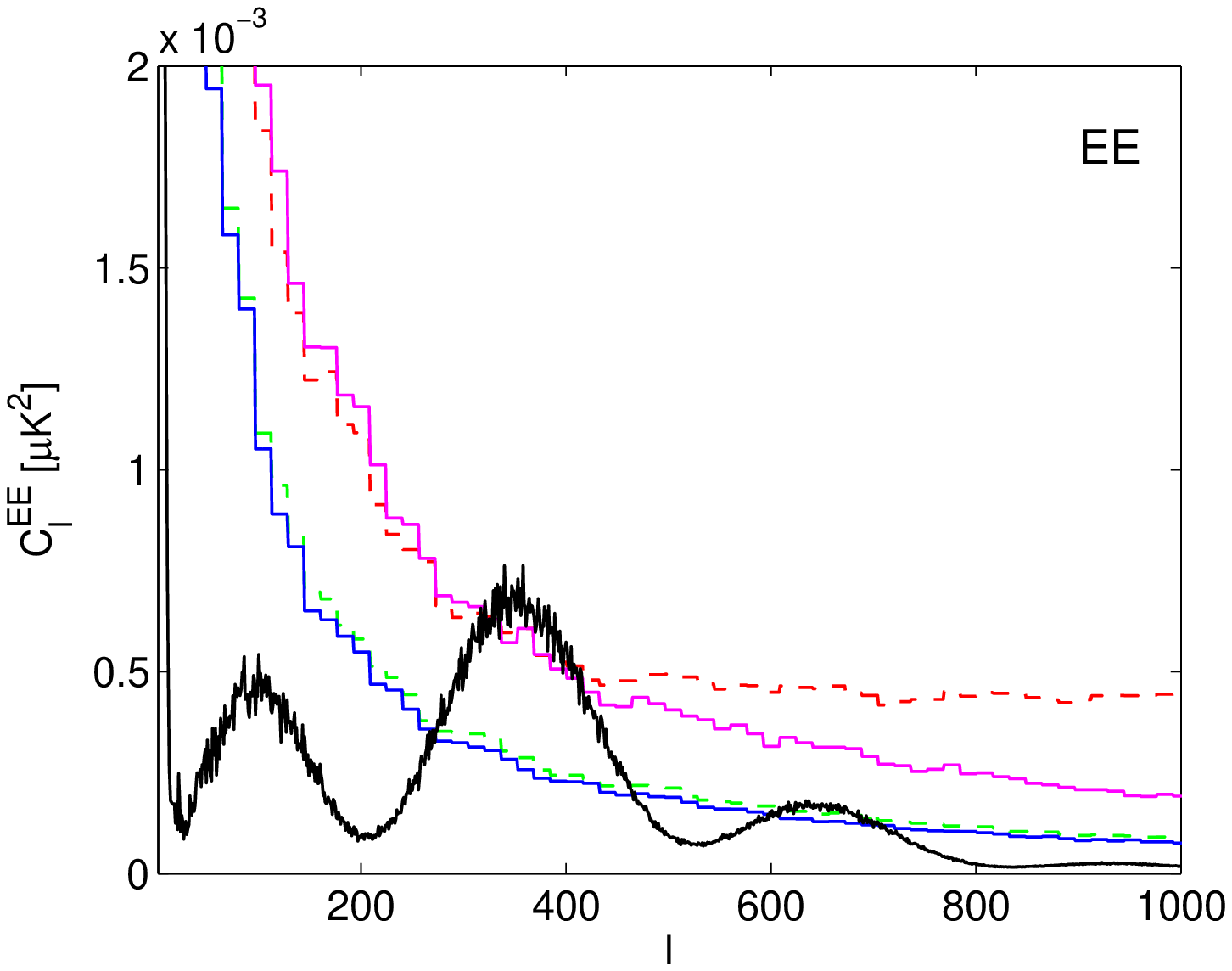}}
 \caption{
  {\bf Noise bias.}
  We show
TT and EE spectra of the CRN (correlated residual noise) map. 
All spectra were computed from {\it nside\_map}=512 maps.
We
compare three baseline lengths: 0.625 s ({\it solid blue} line), 60
s ({\it solid magenta} line), and 3600 s ({\it dashed red}
line). Destriping resolution was {\it nside\_cross}=512 for these three curves.
We show also the effect of a high ({\it nside\_cross}=2048) destriping resolution
({\it green dashed} line.) 
Also shown is the spectrum of the binned CMB map ({\it black
solid} line) and the white noise level ({\it
black dashed} line).  The leftmost panels show the spectra at low multipoles
in logarithmic scale. The rightmost panels show the spectra at high
multipoles in linear scale.
For clarity, the CRN spectra were averaged over 16
adjacent multipoles. 
 } \label{fig:clspec_base}
\end{figure*}


Next we consider the residual noise in the TOD domain. We subtract
the solved baselines from the combined white noise plus $1/f$\ noise
TOD, and compute the spectrum of the residual. We plot the spectrum
for various baseline lengths in Fig.~\ref{fig:todspec1}. We show
results both with and without noise prior. At long baselines (1 min
and above) results with and without noise prior cannot be
distinguished. When we move towards shorter baselines, differences
appear. Without noise prior, residual noise power systematically
decreases with decreasing baseline length. A peak structure appears
at shortest baselines. When a noise prior is used, the spectrum of
residual noise converges towards the spectrum shown by a thick line
in the figure.

It is interesting to note that with a given baseline length,
{\it not} applying a noise prior gives {\it lower} residual noise power in TOD domain.
The situation becomes the opposite when the cleaned TOD is binned into map:
Using a noise prior leads to lower residual noise in map domain.
This is related to the correlation properties of the destriped TOD.
When a noise prior is applied, the residual TOD comes out less strongly correlated
than without noise prior, and its spectrum closer to that of white noise.
The TOD thus averages out more efficiently when binned into a map,
leading to a lower noise level in map domain.

We plot in the same figure an analytical approximation
 \begin{equation}
   P_\mathrm{app}(f) =
     \frac{P_\mathrm{wn}^2}{P_\mathrm{wn}+P_\mathrm{oof}(f)}
     \label{analytical}
 \end{equation}
where $P_\mathrm{wn}$ and $P_\mathrm{oof}$ are the spectra on the
white noise component and the $1/f$\ component, respectively. The
approximation is obtained by setting $\mathbf{F}=\mathbf{I}$ and $\mathbf{Z}=\mathbf{I}$ in the
destriping equation (\ref{linc}). This approximation corresponds to
making the following to assumptions: 1) Assuming that the baseline
length is one sample, so that the baseline vector has the same
length as the TOD itself. This assumption can be expressed as
$\mathbf{F}=\mathbf{I}$. 2) Assuming that there are an infinite number of
observations per pixel, so that we can ignore the pointing matrix
term in the definition of the $\mathbf{Z}$ matrix in Eq.~(\ref{zdef})
when $\mathbf{Z}$ is acting on noise, and
set $\mathbf{Z}=\mathbf{I}$. These are serious approximations, but the analytical
model agrees well with the residual spectrum at high frequencies, at
the short-baseline limit. At low frequencies the approximation is
poor.

The behaviour of the residual noise spectrum in the absence of noise
prior is discussed in detail by \cite{HX2009}.

\subsection{$C_l$ domain}

Finally we study the residual noise in the $C_l$ domain.
We compute the TT and EE angular power spectra of
the CRN map
using the Anafast tool which is part of the HEALPix package.

In Fig.~\ref{fig:clspec_base} we plot the spectrum of the CRN map
for three distinct baseline lengths (1 h, 1 min, and 0.625 s)
together with the spectrum of the binned noiseless map. 
Destriping resolution was {\it nside\_cross}=512.
We show low
and high multipoles separately.
At high multipoles we bin the
spectra over 16 adjacent multipoles, in order to show the
differences more clearly.
The TT plots begin at multipole $l=1$,
the EE plots at $l=2$.

We show also the effect of a high destriping resolution ({\it nside\_cross}=2048).
This case is discussed in more detail in Sect. \ref{sec:resolution}.
All spectra were computed from maps with resolution {\it nside\_map}=512.

We see that a short baseline (0.625 s) gives systematically lower residual noise
than a longer one (1 min or 1 hour) at all but the very lowest multipoles,
though the difference is small as compared to the white noise level.
In the EE spectrum the differences are comparable to the underlying CMB signal.

At the very lowest multipoles the situation is not that clear,
but we remind here that we have one noise realization only,
so the effects seen at lowest multipoles may be somewhat random.

%
%

\section{Destriping resolution}
\label{sec:resolution}

%
%

\begin{figure}
\includegraphics[width=9cm]{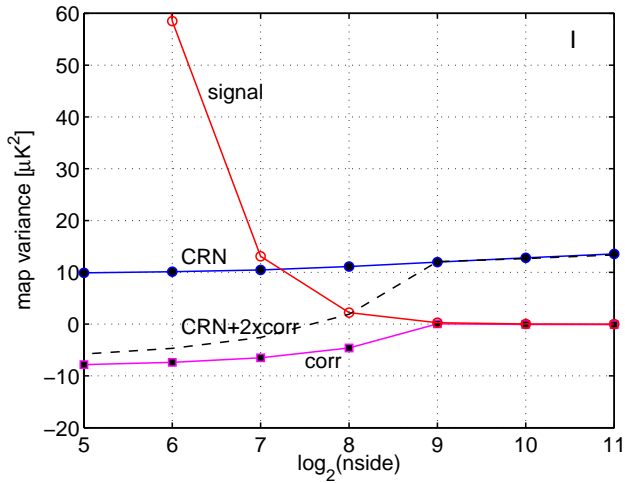}
 \caption{
 {\bf Effect of destriping resolution on residual noise}.
We plot the residual
noise variance against the base-2 logarithm of the {\it nside\_cross}
parameter, for total intensity. We show the
signal error ({\it red, open circles}), CRN ({\it blue, solid
circles}), and covariance between CRN and white noise ({\it purple,
squares}). The {\it black dashed} line shows the sum of the 
CRN variance and twice the covariance
between CRN and white noise. 
 }
\label{fig:nside}
\end{figure}

In this section we study how varying the destriping resolution affects the residual noise.
We also study the effect of applying a galactic mask in the destriping phase.

We vary the destriping resolution {\it nside\_cross}, which defines the crossing points,
but bin the final map at a fixed resolution of {\it nside\_map}=512.
We use baseline length 0.625 s and a noise prior.
The effect of destriping resolution in traditional destriping
was studied earlier by \cite{Maino2002}.

In Fig.~\ref{fig:nside} we show the variance of different residual error components
as a function of destriping resolution.
We destriped the data with 7 distinct HEALPix resolutions: 
(from left to right:
{\it nside\_cross}= 32, 64, 128, 256, 512, 1024, 2048.)
Maps were binned to resolution {\it nside\_map}=512. We plot the residual
noise variance against the base-2 logarithm of the {\it nside\_cross}
parameter, for total intensity. We show the
signal error, CRN, and covariance between CRN and white noise.
The total variance is the sum of these three
and the variance of white noise,
which is 2478 $\mu$K$^2$.
The signal error in the I map increases
rapidly with decreasing resolution, being 189 $\mu$K$^2$ at
{\it nside\_cross}= 32.
Baseline length was 0.625 s.

When the TOD is destriped at a resolution lower than the map
resolution, the binned white noise map and the CRN map, which is
obtained by subtracting the binned white noise map from the total
residual noise, become correlated. For this reason, the variance of
the CRN map alone is not a good figure-of merit. In addition to the
CRN variance we compute the covariance between the CRN map and the
binned white noise map, and plot the it together with the
CRN variance.

We show in the same plot the sum of the CRN variance and twice the
covariance between CRN and binned white noise. The total variance
of the residual noise map is the obtained by adding the variance of
the binned white noise map to this sum. The white noise level is a
constant when {\it nside\_map} is fixed.

%
%
\begin{table}
\caption[a]
{{\bf Reducing signal error.}
We show the signal error and CRN rms ($\mu$K) for total intensity and
for Q polarization, for various parameter settings.
First column is the destriping resolution.
}
\begin{center}
\begin{tabular}{rr|rr|rr}
\hline
\multicolumn{2}{c|}{}  &
\multicolumn{2}{c|}{I}  &
\multicolumn{2}{c}{Q} 
\\
\hline
Nside & mask & Signal & CRN & Signal & CRN \\
\hline
512     & 0\%  & 0.264 & 3.46 & 0.023  & 4.93 \\
512     & 5\%  & 0.076 & 3.50 & 0.021 & 4.97 \\
512     & 10\%  &  0.077 & 3.54 & 0.021 & 5.02\\
1024   & 0\%  & 0.153 & 3.58 & 0.007 &  5.09 \\
1024   & 5\%  & 0.077 & 3.62 & 0.006 & 5.13 \\ 
1024   & 10\%  & 0.078 & 3.66 & 0.006 & 5.15\\ 
2048   & 0\%  & 0.045 & 3.68 & 0.002 & 5.26 \\
2048   & 5\%  & 0.026 & 3.72 & 0.002 & 5.30 \\
2048   & 10\%  & 0.026 & 3.76 & 0.002 & 5.34 \\

\hline
\end{tabular}
\end{center}
\label{tab:resolution}
\end{table}

At low resolutions the sum curve falls below zero. This indicates an
interesting phenomenon. Destriping with a low resolution actually
leads to a residual noise variance {\it below} the white noise
level. This contradicts the often-heard claim that the white noise
level sets a lower limit for the residual noise level that can be
achieved by any map-making method. The drawback is of course that
the residual noise is correlated.

The signal error depends on the resolution much more steeply than
the residual noise. Larger pixels give rise to a larger signal
error, as temperature variations within a pixel become more
important. The signal error for the polarization maps is well below that of
the intensity map, as we showed in Sect. \ref{sec:baseline}.
This reflects the fact that the
polarization signal is weak compared to the total intensity.

The signal error is difficult to remove at later data processing
steps, and hard to model reliably. It may cause undesired artifacts
in the output map. For instance, signal error tends to spread the
image of a strong point source into a line following the scanning
pattern. It is desirable to minimize the signal error at the
map-making level, by destriping at a sufficiently high resolution.

The signal error may be further reduced by masking out the pixels
with highest temperature variation.

Differences in signal error above resolution {\it nside\_cross}=512
are hardly visible in Fig.~\ref{fig:nside}. We study now the high resolution
range ({\it nside\_cross}=512-2048) more closely.

We show our results for the high resolution regime in Table \ref{tab:resolution}.
We show the rms of the signal error and CRN  maps, for total intensity and Q polarization.
Here we have chosen to show the rms, instead of the variance,
in order to have numbers comparable in magnitude.
Maps were binned to a common resolution {\it nside\_map}=512.
Baseline length was 0.625 s in all cases.

We show also the effect of applying a galactic mask
in the destriping phase.
We mask out 5\% or 10\% of the most strongly foreground-dominated
pixels when solving the baseline offsets, but include them again
when binning the cleaned TOD into the final map. 

The reduction in signal error with increasing destriping resolution
continues until our highest resolution {\it nside\_cross}=2048. 
in the same time, the residual noise increases slightly.
Masking
the galaxy reduces the signal error in the intensity map
significantly. The difference between 5\% and 10\%
masks is small.
Because our simulation data set did not include polarized foregrounds,
masking the galaxy has little effect on the signal error in the polarization map.

%
%
\begin{figure}
\includegraphics[width=9cm]{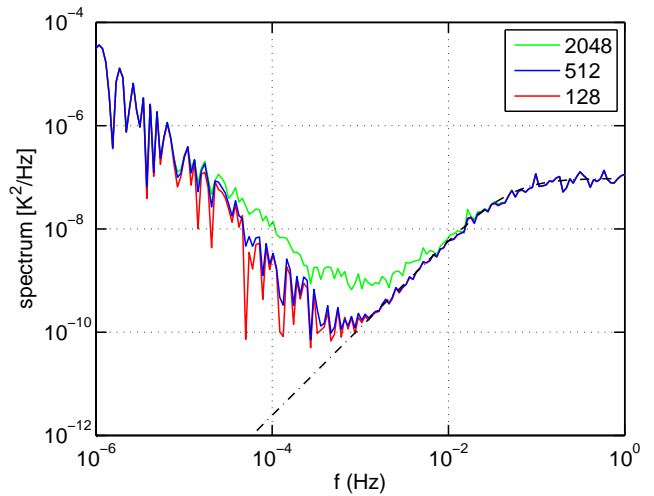}
\caption{ 
{\bf Effect of destriping resolution on the residual noise spectrum}. 
We plot the spectrum  of residual noise TOD, obtained by
subtracting the solved baselines from the original noise TOD. The
baselines were solved at resolution (from top down) {\it nside}=2048
({\it green}), 512 ({\it blue}) or 128 ({\it red}). 
The
{\it black dashed} line presents the analytical approximation given
by Eq. (\ref{analytical}). } \label{fig:todspec_nside}
\end{figure}


As in the previous section, where we studied the effect of baseline
length, we study the noise residual also in the time domain and in
the $C_l$ domain.
In Fig.~\ref{fig:todspec_nside} we plot the spectrum of the residual
noise TOD for three destriping resolutions, {\it nside\_cross} = 2048, 512,
128. Baseline length was again 0.625 s in all cases. We show also the
analytical approximation given by Eq. (\ref{analytical}).

A high destriping resolution clearly leaves more noise in the TOD at
intermediate scales. 

Going from resolution {\it nside\_cross}=512 to {\it nside\_cross}=128 has little
effect on the spectrum above 1 mHz, but brings the spectrum down at
lower frequencies. At the same time, the peaky structure in the
spectrum becomes more pronounced.

Note that the spectrum of the residual noise TOD is only dependent
on the destriping resolution {\it nside\_cross}, and has nothing to do with
the map resolution. Thus the dependence or independence of CNR and
white noise in the map domain plays no role here.

Though the difference between {\it nside\_cross}=512 and {\it nside\_cross}=2048
shows clearly in the residual noise spectrum,
the effect is less dramatic in the angular power spectrum of the CRN map.
We compare the two cases in Fig. ~\ref{fig:clspec_base}.
The higher destriping resolution leaves slightly more noise
at nearly all multipoles,
but the effect is small compared with the effect of the baseline length.

%
%
\section{Conclusions}

We have  presented an update of the {\small MADAM}\  map-making method and
applied it to simulated Planck-like data. We produced maps of total
intensity and Q and U polarization.

The {\small MADAM}\ algorithm is tuned by selecting values for a set of input parameters.
We have studied the effect of baseline length and destriping resolution
on residual error.

{\small MADAM}\ differs from traditional destripers in that it uses a noise
prior, which allows to extend the method to very short baselines,
which model the noise better. The noise prior has little effect on
the results at baseline lengths longer than the inverse of the knee frequency,
but becomes important at short baselines.

We varied the baseline length from 0.1 seconds to over one hour.
We obtained best results when the baseline length was below one second.
Our simulations assumed a knee frequency of 50 mHz.

For a Planck-like scanning strategy, long baselines, longer
than the spin period, up to the repointing period, allow a
significant reduction in the computer memory requirement. For these
long baselines, the level of residual error depends strongly on the
baseline length. ÊThis is related to the sky scanning pattern.
Baseline lengths which are an integer fraction of the repointing
period, are strongly favored over other baseline lengths.

The strength of {\small MADAM} is in its flexibility in the choice of baseline length.
A long baseline gives a quick-and-dirty map,
while a short baseline gives a high accuracy map.
Some guidelines for the selection of baseline length can be given.
When working with a new data set, it us usually safe to start with an intermediate
baseline length (1 min or equal to the scanning period)
and to do the destriping without noise prior.
If the noise properties of the data are well known,
and if computational resources allow,
the best accuracy is obtained with a noise prior and with a very short baseline,
which models the non-white part of the noise.
A long baseline (longer than the scanning period) allows to significantly
reduce the memory requirement, but then it is important to keep in mind
that the accuracy of the resulting map may depend strongly on
the scanning pattern. The split-mode offers another means of reducing
 the memory requirement.

{\small MADAM}\ allows to build the output map at a resolution different
from the destriping resolution. We found that if destriping is
performed at a resolution lower than the map resolution, the
residual noise power may fall below the white noise level. 
Real maps,
however, are unlikely to be build this way, since the control of
signal error requires that destriping is done at high resolution.

The signal error can be reduced by destriping at a high resolution,
and by applying a galactic mask in the destriping phase.

We studied the properties of residual error in terms of
the variance of the residual map, spectrum of residual noise,
and the angular power spectrum of the residual noise map.
All these demonstrate different aspects of the residual error.


\begin{acknowledgements}

The work reported in this paper was done as part of the CTP Working Group of
the {\sc Planck} Consortia. {\sc Planck} is a mission of the
European Space Agency. This work was supported by the Academy of
Finland grants 205800, 213984, 214598, 121703, and 121962. RK is
supported by the Jenny and Antti Wihuri Foundation. HKS thanks
Waldemar von Frenckells stiftelse, HKS and TP thank the Magnus
Ehrnrooth Foundation, and EK and TP thank the V\"{a}is\"{a}l\"{a}
Foundation for financial support. This work was supported by the
European Union through the Marie Curie Research and Training Network
``UniverseNet'' (MRTN-CT-2006-035863). We thank CSC (Finland) for
computational resources.   We acknowledge use of the {\sc CAMB} code
for the computation of the theoretical CMB angular power spectrum.
This work has made use of the {\sc Planck} satellite simulation
package (level S), which is assembled by the Max Planck Institute
for Astrophysics {\sc Planck} Analysis Centre (MPAC). Some of the
results in this paper have been derived using the HEALPix package
\cite{Gorski05}. We acknowledge the use of the Planck Sky Model,
developed by the Component Separation Working Group (WG2) of the
Planck Collaboration.

\end{acknowledgements}

%

\bibliographystyle{aa}


%
\end{document}